\documentclass[aasmacros]{aa}

\usepackage{amsmath}
\usepackage{wasysym}
\usepackage{marvosym}
\usepackage{txfonts}
\usepackage{mathtools}
\usepackage{placeins}
\usepackage{siunitx}
\usepackage{enumitem}
\DeclareSIUnit[]\astronomicalunit{\text{au}}
\DeclareSIUnit\year{yr}
\usepackage[symbol]{footmisc}
\usepackage[T1]{fontenc}
\usepackage{natbib}
\bibpunct{(}{)}{;}{a}{}{,} 

\usepackage{epstopdf}
\usepackage{graphicx}
\pdfoutput=1
\usepackage[pdftex,dvipsnames]{color,xcolor}
\usepackage{subcaption}

\usepackage{scalerel}
\usepackage{tikz}
\usetikzlibrary{svg.path}
\usepackage{multirow}
\usepackage{pdflscape}
\usepackage{afterpage}
\usepackage{soul}
\usepackage[export]{adjustbox}

\definecolor{orcidlogocol}{HTML}{A6CE39}
\tikzset{
  orcidlogo/.pic={
    \fill[orcidlogocol] svg{M256,128c0,70.7-57.3,128-128,128C57.3,256,0,198.7,0,128C0,57.3,57.3,0,128,0C198.7,0,256,57.3,256,128z};
    \fill[white] svg{M86.3,186.2H70.9V79.1h15.4v48.4V186.2z}
                 svg{M108.9,79.1h41.6c39.6,0,57,28.3,57,53.6c0,27.5-21.5,53.6-56.8,53.6h-41.8V79.1z M124.3,172.4h24.5c34.9,0,42.9-26.5,42.9-39.7c0-21.5-13.7-39.7-43.7-39.7h-23.7V172.4z}
                 svg{M88.7,56.8c0,5.5-4.5,10.1-10.1,10.1c-5.6,0-10.1-4.6-10.1-10.1c0-5.6,4.5-10.1,10.1-10.1C84.2,46.7,88.7,51.3,88.7,56.8z};
  }
}

\newcommand\orcidicono[1]{\href{https://orcid.org/0000-0001-6693-7910/#1}{\mbox{\scalerel*{
\begin{tikzpicture}[yscale=-1,transform shape]
\pic{orcidlogo};
\end{tikzpicture}
}{|}}}}

\newcommand\orcidiconc[1]{\href{https://orcid.org/0000-0002-1013-2811/#1}{\mbox{\scalerel*{
\begin{tikzpicture}[yscale=-1,transform shape]
\pic{orcidlogo};
\end{tikzpicture}
}{|}}}}

\newcommand\orcidicona[1]{\href{https://orcid.org/0000-0002-8811-1914/#1}{\mbox{\scalerel*{
\begin{tikzpicture}[yscale=-1,transform shape]
\pic{orcidlogo};
\end{tikzpicture}
}{|}}}}

\newcommand\orcidiconr[1]{\href{https://orcid.org/0000-0001-5555-2652/#1}{\mbox{\scalerel*{
\begin{tikzpicture}[yscale=-1,transform shape]
\pic{orcidlogo};
\end{tikzpicture}
}{|}}}}

\usepackage[draft=False, colorlinks=true, citecolor=blue, urlcolor=blue, linkcolor=red, linktoc=all]{hyperref}

\usepackage{pifont}
\def\Haek{\ding{51}}

\def\mearth{M_\oplus}

\def\msun{M_\odot}

\def\f1{f_{\rm I}}
\def\mstar{M_*}

\def\beq{\begin{equation}}
\def\eeq{\end{equation}}

\def\t2{\tau_{\rm II}}

\def\sigmas0{\Sigma_{\rm s,0}}
\def\mj{M_{\textrm{\tiny \jupiter }}}

\def\s0{S_0}

\def\apjl{ApJ}                
\def\apjs{ApJS}               
\def\ao{Appl.Optics}          
\def\aap{A\&A}                

\def\({\left(}
\def\){\right)}
\def\<{\left<}
\def\>{\right>}

\defcitealias{2021A&A...645A..43S}{S21}
\defcitealias{2022A&A...664A.138S}{S22}
\defcitealias{2023A&A...669A..31S}{S23}
\defcitealias{2018MNRAS.475.5618B}{B18}
\defcitealias{Kimura2016}{K16}
\defcitealias{Schib2025}{Paper~I}

\begin{document}

\title{DIPSY: A new \textbf{D}isc \textbf{I}nstability \textbf{P}opulation \textbf{SY}nthesis\footnotemark}
\subtitle{II. The Populations of Companions Formed Through Disc Instability}

    \author{O. Schib\inst{\ref{WP}\orcidicono{}}
          \and
          C.~Mordasini\inst{\ref{WP},\ref{CSH}\orcidiconc{}}
          \and
          A.~Emsenhuber\inst{\ref{WP}\orcidicona{}}
          \and
          R.~Helled\inst{\ref{uzh}\orcidiconr{}}
          }

   \institute{
Space Research and Planetary Sciences, Physics Institute, University of Bern, Gesellschaftsstrasse 6, 3012 Bern, Switzerland\\
\email{oliver.schib@unibe.ch}
\label{WP}
\and
Center for Space and Habitability, University of Bern, Gesellschaftsstrasse 6, 3012 Bern, Switzerland\\
\label{CSH}
\and
Department of Astrophysics,
Universit\"at Z\"urich, Winterthurerstrasse~190, 8057~Z\"urich, Switzerland\\
\label{uzh}
             }

\date{Received July 4, 2025 / Accepted September 1, 2025}

\abstract
{Disc instability (DI) might provide an explanation for the formation of some observed exoplanets. 
At the same time, our understanding of this top-down formation mechanism remains limited. 
Existing studies have made strong simplifications, and the predicted population is poorly known.}
{We aim at overcoming several limitations and produce a more advanced synthetic population of companions formed via DI that can be used for quantitative statistical comparisons with observations, and to make predictions for future surveys.}
{We applied the global end-to-end model described in Paper~I of this series to perform a population synthesis of companions formed via DI.
By using initial conditions compatible with both observations and hydrodynamical simulations, and by studying a large range of primary masses (\SIrange{0.05}{5}{\msun}), we can provide quantitative predictions of the outcome of DI.}
{In the baseline population, we find that $\sim$\SI{10}{\%} of the discs fragment, and about half of these end up with a surviving companion after \SI{100}{Myr}.
Based on their mass, 75\% of the companions are in the brown dwarf regime, \SI{15}{\%} are low-mass stars, and \SI{10}{\%} planets.
At distances larger than $\sim$100 AU, DI produces planetary-mass companions on a low percent level.
Inside of 100 AU, however, planetary-mass companions are very rare (low per mill level). The average companion mass is $\sim$30 $\mj$ scaling weakly with stellar mass. Very few companions of all masses reside inside of 10 AU; outside this distance, the distribution is approximately flat in log-space. Eccentricities and inclinations are significant, with averages of 0.4 and 40$^{\circ}$. In systems with surviving companions, there is either one (80\%) or two (20\%) companions. The fraction of surviving synthetic brown dwarfs is consistent with observations, while that of planets is lower than observed.
Most of the initial fragments do not survive on a \unit{Myr} timescale; they either collide with other fragments or are ejected, resulting in a population of free-floating objects (about 1-2 per star). 
We also quantify several variant populations to critically assess some of our assumptions used in the baseline population.} 
{DI appears to be a key mechanism in the formation of distant companions with masses ranging from low-mass stars down to the planetary regime, contributing, however, only marginally to planetary mass objects inside of 100 AU.
Our results are sensitive to a number of physical processes, which are not completely understood.
Two of them, gas accretion and clump-clump collisions, are particularly important and need to be investigated further. 
Magnetic fields and heavy-element accretion have not been considered in our study, although they are also expected to affect the inferred population. 
We suggest acknowledging the importance of the gravito-turbulent phase, which most protoplanetary discs experience. Exploring hybrid DI -- core accretion scenarios, and quantitative comparisons of theory and observations will improve our understanding of star and planet formation.}

\maketitle

\section{Introduction}\label{sec:intro}

\footnotetext{*Companion data is available for download at \url{https://www.space.unibe.ch/research/research_groups/theoretical_astrophysics_and_planetary_science_taps/numerical_data/index_eng.html}}
Planetary population synthesis is a method used to put different theories of planet formation to the quantitative observational test by statistically comparing their predictions with the observed exoplanet population \citep[e.g.][]{2004ApJ...604..388I,2009A&A...501.1139M}. It has become possible thanks to the fact that over the last three decades since \citet{1995Natur.378..355M}, large surveys employing different observational techniques have discovered thousands of exoplanets as well as sub-stellar companions. Important examples of radial velocity surveys include the HARPS survey \citep{Mayor2011} and the California Legacy survey \citep{Rosenthal2021}; of the transit method, the Kepler satellite \citep{Borucki2010,Petigura2018}; direct imaging surveys such as GPIES \citep{Nielsen2019}, SPHERE SHINE \citep{2021A&A...651A..72V}, and BEAST \citep{2021A&A...646A.164J}; and finally, microlensing \citep{Suzuki2016}.

Observations of planets as they form remain challenging, and to date, only a handful of bona fide examples are available \citep{Keppler2018}.
As a result, the process of planet formation cannot be directly constrained from observations.
The limited knowledge about most individual exoplanets (such as some orbital elements and a mass or radius) further affects our ability to construct planet formation models, because they cannot be easily confirmed or rejected through these approaches.

On the other hand, statistical approaches can be used to better constrain planet formation theory.
They make use of the large number of observations of young stellar objects (YSOs) and, in particular, their discs \citep{Andrews2020} in which planets and other companions form.
The observations of discs and YSOs mark the initial and boundary conditions for formation.
Observed (exo)planetary systems with their statistical properties \citep{Udry2007} represent the final outcome of the formation process.
This is the principle of planetary population synthesis \citep{2014prpl.conf..691B,Burn2024}, which tries to find a theory that can best reproduce the observed demographics of planetary systems.

The most important element of the population synthesis approach is a global formation model \citep{mordasinimolliere2014,2022ASSL..466....3R}, which can produce synthetic populations.
It is based on an underlying formation paradigm and combines the results of many models of specific physical processes such as disc formation and evolution, gas accretion, collisions, and orbital migration.

Two fundamentally different planet formation paradigms have been proposed. Disc instability (DI), also known as gravitational instability (GI), was the first proposed for the formation of the Solar System \citep{1951PNAS...37....1K}.
It was later mostly superseded by the core accretion (CA) model \citep{safronov1972,Pollack1996,Alibert2005} as the main formation pathway of Solar System planets as well as observed exoplanets. However, the DI model is still a valid alternative in some cases and has remained the favoured explanation for the formation of massive giant planets at large separation.  
The formation of these systems is hard to explain by CA. A famous example of such a system might be the HR~8799 system \citep{2008Sci...322.1348M,2010Natur.468.1080M}.

While not studied as much in detail as CA, a number of important results concerning the outcome of DI exist.
A fundamental prediction is that DI should, in particular, form brown dwarfs (BDs).
For example, \citet{Stamatellos2009} performed radiation hydrodynamics simulations of fragmenting discs and found that the most likely outcome is companions in the BD mass range with high eccentricity and inclination, although stellar and planetary-mass companions were also formed.
\citet{Kratter2010} studied fragmentation and the subsequent evolution of fragments and investigated whether wide companions such as those in the HR~8799 system could form through DI.
They explored the influence of the initial fragment mass, gas accretion, and gap formation on the final masses and found that the formation of an object with a final mass in the planetary regime is rather unlikely.
Instead, masses in the BD or stellar regime were found to be the most probable outcome.

\citet{helledbodenheimer2010a} studied the composition of companions formed in DI.
The influence of a number of factors, most notably the pre-collapse timescale, was considered in planetesimal capture through clumps.
The authors found that, if formed in DI, the HR~8799 planets would have low metallicities.

 \citet{2012ApJ...746..110Z} performed 2D hydrodynamical simulations of self-gravitating discs including infall.
They studied the accretion of gas and the migration of the clumps formed, and concluded that many are lost by Type~I migration to primary or tidal destruction, although gap formation is possible in some cases due to gas accretion.

\citet{2010Icar..207..509B} studied the initial conditions for clumps formed in DI.
They provided an analytic estimate for the initial clump mass that places this quantity in the gas giant regime.
This estimate was confirmed in radiation hydrodynamics simulations.
It was found that clumps could form a heavy-element core and have a tidally stripped envelope, thus potentially leading to the formation of rocky planets.

\citet{Baehr2019} performed 3D self-gravitating shearing sheet simulations in order to study solid accretion by clumps.
They find that the formation of cores of up to several Earth masses is possible.
The process will depend on the metallicity of the disc. Core formation leads to a solid depletion of the clump's atmosphere, whereas if no core forms, a high metallicity atmosphere is expected.
For a more thorough overview, we refer to the reviews in \citet{2007prpl.conf..607D,2010fee..book...71M,2014prpl.conf..643H,2016ARA&A..54..271K}.

The population of companions formed via DI is currently unknown, and it remains unclear whether some observed companions formed in this way.
In order to answer this question, it is necessary to perform a population synthesis study that links observationally constrained initial conditions to a population of surviving companions, considering all relevant physical processes.
Various previous studies have aimed at establishing a population synthesis model for DI, but they have limitations (see Sect.~\ref{sec:GIPop} for discussion).
A key limitation is that previous studies did not include the earliest formation phase where the star-and-disc system forms during infall.
This phase is crucial because this is when the disc is most massive and prone to fragmentation.
Another key limitation of existing studies is that the initial conditions are not statistically linked to observations.
Especially, the fraction of systems that fragment is important to determine.
This fragmentation fraction directly translates into the fraction of systems that can potentially host a companion formed through DI.
In this paper, we present a comprehensive population synthesis model that aims to overcome these issues.
We applied the model described in \citet{Schib2025} (Paper~I) with initial conditions informed by observed quantities such as stellar masses and disc sizes, as well as results from hydrodynamic simulations.
This allows us to make predictions about the properties of formed systems and their occurrence rates.
For example, the fraction of companions in a given interval of mass and separation can be predicted as a function of the primary mass.
Such predictions enable a quantitative comparison to observational surveys \citep[e.g.][]{Nielsen2019}.
Furthermore, it is possible to estimate the contribution of DI to the observed exoplanet population and guide future observations as facilities improve \citep[e.g.][]{Lagrange2025}.

This paper is organised as follows:
In Sect.~\ref{sec:pop} we give a brief overview of the population synthesis method with a focus on existing projects in the DI paradigm.
Section~\ref{sec:model} briefly summarises the model.
The initial conditions for the population synthesis are described in Sect.~\ref{sec:init}.
In Sect.~\ref{sec:res} we present the results.
Section~\ref{sec:add} is dedicated to the variant populations we performed in order to study a number of parameters.
Section~\ref{sec:comp} compares results from previous studies.
In Sect.~\ref{sec:compo}, we compare with observations.
Section~\ref{sec:Discussion} contains a discussion of our results and limitations of the model.
We summarise and conclude in Sect.~\ref{sec:Conclusions}.

\section{Planetary population synthesis}\label{sec:pop}

\subsection{The population synthesis method}\label{sec:method}

Planetary population synthesis requires several ingredients \citep{2014prpl.conf..691B,Emsenhuber2023,Burn2024}, including demographics of YSOs and their discs for the initial conditions, a formation model to predict an exoplanet population, and exoplanet demographics to compare with the model results. The model must meet specific requirements: it must be global (i.e. including all the known relevant physical processes and them being coupled), end-to-end (i.e. going from YSOs to mature systems, predicting as many directly observable properties as possible), but still remain relatively fast computationally such that a full population with thousands of systems can be computed.

In the CA paradigm, these criteria are met by a number of existing studies, where models have reached a significant level of complexity. Thus, there are detailed predictions about the expected population of CA planets. Conversely, only a few GI population synthesis projects exist. They are less comprehensive in comparison to their CA counterparts and have received less attention in the literature. This reflects that global models can only be constructed if sufficient specialised models on the individual relevant physical processes exist.
Below, we briefly review some of the existing CA projects, followed by a more detailed description of published GI-population studies.

\subsection{CA population synthesis models}\label{sec:CApop}
The era of planetary population synthesis began with the work of \citet{2004ApJ...604..388I}. The authors developed a simple global model of planet formation and used it to study the mass-semi-major axis distribution of planets predicted by CA.
They included stars of different masses and metallicities, but neglected Type~I orbital migration.
Since then, CA population synthesis and global models have been improved and include more processes such as the interior structure, N-body interactions between planets, and the long-term evolution of planets. In addition, more detailed comparisons with the increasingly large observational surveys have been performed.
Key studies include \citet{idalin2008,idalin2008c,2008Sci...321..814T,2009A&A...501.1139M,mordasinialibert2009b,miguelguilera2011,2011MNRAS.417..314M,alibertmordasini2011c,2012MNRAS.419.2737H,alibertcarron2013,2010ApJ...719..810I,idalin2013,Coleman2014,2016MNRAS.457.2480C} and references therein. 
The most extensive population synthesis project to date is the Bern Model \citep{2009A&A...501.1139M,mordasinialibert2009b,alibertcarron2013,2021A&A...656A..70E,2021A&A...656A..71S,2021A&A...656A..72B,2021A&A...656A..73S,2021A&A...656A..74M,Emsenhuber2023}.
Further details on planetary population synthesis models in the CA paradigm can be found in several review papers, including \citet{2014prpl.conf..691B,2018haex.bookE.143M,Drazkowska2023,Emsenhuber2023,Burn2024}.

\subsection{DI population synthesis models}\label{sec:GIPop}
Although GI has not been studied as extensively as CA, a few DI population synthesis models have been published. Two key models are the `Forgan \& Rice model' \citep{2013MNRAS.432.3168F,2018MNRAS.474.5036F} and the `Nayakshin model' \citep{2015MNRAS.454...64N,2015MNRAS.452.1654N,2016MNRAS.461.3194N}.
The existing studies have rather different foci, and they also differ significantly in the choice of initial conditions and the physical processes that are included, as we discuss below.
A table that compares the most important features of the different projects is presented in Appendix~\ref{app:comp}.

\subsubsection{Forgan \& Rice model}\label{sec:forganrice}
D.~Forgan and K.~Rice published a series of papers on self-gravitating discs and fragmentation. Their first population synthesis model was presented in \citet{2013MNRAS.432.3168F}.
Their population focused on solar mass stars and used pre-evolved discs to study the outcome of disc fragmentation.
Fragments with an initial fragment mass $M_\mathrm{J,FR}$ (Sect.~\ref{sec:namj}) were inserted at random locations within the disc and were allowed to migrate.
The fragments could migrate and open gaps. Grain growth and settling were also included, while gas accretion was not considered. 
The authors found that the majority of surviving companions are BDs, and about \SI{40}{\%} form heavy-element cores.
The formation of terrestrial planets was found to be very unlikely.
In \citet{2018MNRAS.474.5036F}, several adjustments to the migration prescription were made, and more importantly, gravitational interactions between fragments were included using N-body simulations (though mergers were not considered). 
It was found that gravitational interactions dominate the orbital evolution, leading again to a population dominated by BD, but with a much wider spread in semi-major axes.
Some of the fragments with an evolved grain content were found to be disrupted, possibly leading to the formation of planetesimal belts.
The authors also predicted a population of free-floating objects due to ejections.

\subsubsection{Nayakshin model}\label{sec:nayak}
S.\,Nayakshin and collaborators presented a population synthesis study focused on tidal downsizing, which is the fragmentation of the gaseous disc to form bound clumps followed by solid accretion by the clumps, core formation, and tidal stripping of the gas envelope (\citealt{2015MNRAS.454...64N,2015MNRAS.452.1654N,2016MNRAS.461.3194N}).
These models focused on solar mass stars and used one companion per system.
Gas accretion was not considered, but instead, fragments could accrete pebbles.
Planet-disc interaction was modelled by means of a simple Type~I migration timescale, and by the impulse approximation in the gap-opening regime.
The authors found a population of companions with masses from \SIrange{e-3}{2e1}{\mj} occupying the region from \SIrange{e-1}{e2}{au}.
The population includes rocky planets as well as gas giants with a range of bulk metallicities.\\

\subsubsection{Other studies}\label{sec:otherGI}
Here we briefly mention a few other studies that aimed to understand the planetary population predicted by the DI paradigm.
\citet{2011MNRAS.416.1971B} performed 2D hydrodynamic simulations of gravito-turbulent discs with prescribed cooling.
They studied the migration of a single clump around a \SI{1}{\msun} star and found it to migrate inwards very quickly ($\sim$~\SI{10}{kyr}) without forming a gap, independent of clump mass. 
\citet{2020MNRAS.496.1598R} studied a very similar setting, but included a prescription for cooling that has a longer timescale in the inner disc.
It was found that this more realistic cooling allows the clumps to open a gap in the disc.
As a result, clumps are saved from catastrophic inward migration.

\citet{2018ApJ...854..112M} explored the migration of single clumps in static (after fragmentation) discs around predominantly solar mass stars.
They investigated different gap opening criteria, initial clump masses, and clump density profiles.
Gas accretion and tidal mass loss were also considered.
They find that surviving clumps on wide orbits are rare, and that the inferred population depends strongly on the model assumptions.

\citet{2019MNRAS.488.4873H} performed simulations of migrating clumps formed in DI, aiming at constraining their migration using constraints from radial velocity surveys of massive giants inside \SI{5}{au}.
They simulated single evolving clumps around solar type stars, neglecting accretion of gas or solids.
Thermal irradiation and tidal disruption were considered. 
The authors found that tens of percent of systems need to fragment, and clumps need to migrate quickly to the inner disc, in order for their constraints to be met.

\citet{2020ApJ...904...55J} conducted a population synthesis study using a 1D model with a simple infall model.
They varied the properties of the parent molecular cloud core (MCC; temperature, mass, and rotation) and investigated the influence on the initial fragment mass as well as the location and time of fragmentation.
Applying the Toomre mass \citep{Nelson2006}, they inferred initial masses typically \SIrange{3}{35}{\mj} and fragmentation locations between \SIrange{20}{200}{au}.

\section{Model}\label{sec:model}
The model used in this work was developed, applied, and extended in \citet{2021A&A...645A..43S}~(S21), \citet{2022A&A...664A.138S}~(S22), and \citet{2023A&A...669A..31S}~(S23). The updated version used in the present work is described in detail in \citetalias{Schib2025}.
Thus, we only give a brief summary here.

The model describes the fragmentation of discs around single stars.
In each simulation, the star-and-disc system is initialised at a very small mass and then grows from infalling material from the MCC.
To this end, the 1D viscous evolution equation for the disc's radial structure is solved numerically, yielding the disc's surface density \citep{1952ZNatA...7...87L,1974MNRAS.168..603L}. The disc's vertical structure and temperature are calculated based on a thermal balance criterion, including the effects of accretion heating, viscous heating, stellar irradiation, and radiative cooling (\citealt{1994ApJ...421..640N,2005A&A...442..703H},\citetalias{2023A&A...669A..31S}).

The infalling material is added to the disc, realised as a source term in the equation.
Matter accreting across the fixed inner edge of the disc is added to the star.
Clumps (young fragments, bound blobs of gas) emerge when the conditions for fragmentation are satisfied.
Their initial mass is assumed to be $M_\mathrm{F}$  (\citealt{2010Icar..207..509B}, Eq.~35 in \citetalias{Schib2025}).
The corresponding gas is removed from the disc.
Clumps evolve according to pre-calculated evolution tracks of isolated 1D objects, corrected by disc irradiation \citep{2006Icar..185...64H,2008Icar..195..863H,2012ApJ...756...90V,vazankovetz2013,2015ApJ...803...32V,2019MNRAS.488.4873H}.
The evolution tracks include the second collapse of the clumps due to the dissociation of molecular hydrogen \citep{1974Icar...23..319B}.
If this happens, the clump sizes shrink by several orders of magnitude, and the objects become compact gas spheres. These  are gravitationally bound and can end up as giant planets, BDs, or stars.

Clumps can gain mass by gas accretion and collisions and lose mass due to stellar tides.
They can also be completely disintegrated by tidal or thermal disruption.
Any mass that is lost by the clump is returned to the disc, where the exchange of mass between companions and the disc is modelled with a source/sink term in the evolution equation to ensure the conservation of mass.

Companions migrate and open gaps via two-way exchange of angular momentum, which is realised through torque densities in the disc surface density evolution equation \citepalias{2022A&A...664A.138S}.
Damping is calculated based on dynamical friction \citep{2020MNRAS.494.5666I}, and the gravitational interaction between companions is modelled using N-body integration \citep{Chambers1999}.
Clumps can become unbound and leave the planetary system, or be accreted by the central star.
Companions can also collide, and in this work, we assume that collisions lead to perfect merging (see discussion in Sect \ref{sec:disc_coll}).

The systems were evolved until the disc disappeared.
If any companions remained at this point, we continued the N-body integration up to \SI{100}{Myr}.

The model thus includes a significant number of physical processes that interact and feed back on each other in multiple ways. To still be able to simulate a high number of systems from t=0 to 100 Myr, the model is low-dimensional, as typical for global models: the disc is 1+1D (radial and vertical) while the clumps/companions are 1D spherically symmetric. Only the N-body is fully 3D. 

\section{Initial conditions}\label{sec:init}
To perform a population synthesis, the initial conditions for the simulations must be defined. This includes, on the one hand, parameters that are fixed for all simulations, and on the other hand, Monte Carlo variables that are drawn from given probability distributions. 
Table~\ref{tab:param} lists the most important fixed parameters (top) as well as the Monte Carlo variables (bottom).
For the latter, the mean value is given.
The quantities are explained in the following.
\begin{table}
\caption[]{Quantities used in the baseline population DIPSY\nobreakdash-0.}
\centering
\begin{tabular}{ccccccc}  
\hline\hline
\begin{tabular}[c]{@{}c@{}}Parameter\end{tabular} &
\begin{tabular}[c]{@{}c@{}}value\end{tabular}\\
\hline
$\alpha_\mathrm{BG}$ & \num{e-2} \\
$S_\mathrm{wind}$ & \SI{2.8e-8}{g.cm^{-1}.yr^{-1}} \\
\hline
$M_{*,init}$ & \SI{0.05}{\msun} \\
$M_\mathrm{d,init}$ & \SI{0.015}{\msun} \\
$R_\mathrm{d,init}$ & \SI{13}{au} \\
$\dot{M_\mathrm{in}}$ & \SI{2.7e-5}{\msun.yr^{-1}} \\
$t_\mathrm{in}$ & \SI{30.1e3}{yr} \\
$e_\mathrm{init}$ & \num{0.15} \\
\hline
\end{tabular}
\tablefoot{Fixed quantities are at the top, while Monte Carlo quantities are at the bottom.
For the Monte Carlo variables, the mean value is given.
See text Sect.~\ref{sec:init} for further explanations.}
\label{tab:param}
\end{table}

\subsection{Overall setup}
In planetary population synthesis for the CA paradigm, setting initial conditions typically means defining the initial (gas) disc mass and the (fixed) stellar mass. The outer radius and profile of the gas disc are then defined based on observational constraints and theoretical considerations. During the simulation, the disc's mass can only decrease, and the stellar mass remains constant (for example, Sect.~3 in \citealt{2021A&A...656A..70E}).

Since for DI we want to study the earliest stages of planet formation when discs are most massive, we needed a different approach.
Our star-and-disc systems are initialised very early, at \SI{1}{\kilo\year} after the formation of the central protostar.
The masses of the star and the disc at this time are of the order of \SI{0.01}{\msun}.
This is very low compared to their maximum value reached during the evolution.
The large majority of the mass is added to the system by infall, where all the infalling material is added to the disc.
The star's mass is increasing as the disc gas accretes onto the star across the inner edge. During the simulation, the masses of the star and disc increase significantly, in contrast to CA simulations. 
Knowledge of infall rates at the earliest moments of star formation is therefore required. 
In addition, we required some information about the sizes of young discs in order to identify the locations where gas infalls onto the disc. 

\subsection{Initial star and disc masses and infall rates}\label{sec:massacc}
The stellar masses, disc masses, and disc sizes are determined by the source term for the infall.
Unfortunately, these quantities and their time evolution are difficult to observe directly and therefore we built upon the numerical results from \citet{2018MNRAS.475.5618B} where a hydrodynamic simulation of star formation \citep{Bate2012} was analysed in detail with respect to the discs formed around the forming stars.
The original calculation consists of a \SI{500}{\msun} molecular cloud with a radius of $\approx$~\SI{8e4}{au} and uses \num{3.5e7} smoothed particle hydrodynamics (SPH) particles.
During the simulation, \num{183} protostars (represented by sink particles) are formed, and the discs around these protostars are analysed.
An important result of the study is the chaotic nature of the star formation process.
It shows that the ideal picture of a disc around a single star in isolation is over-simplified and rather rare. 
Instead, most discs are truncated in star-disc interactions, multiple systems form, and discs can be destroyed and formed again through accretion.
It is therefore challenging to extract the required data. 
For simplicity, we selected 35 single (i.e. not bound to another star) protostars from the SPH simulation.
The selection criterion was that these systems are not affected by other protostars for a time span of $\approx$~\SI{10}{kyr}.

We used the growth rates of these systems (protostar + disc) in this early phase as infall rates for our simulations.
The indices of the systems we chose are given in Appendix~E of \citetalias{2021A&A...645A..43S}.
We assumed that the stellar mass, disc mass, and infall rate extracted this way are correlated, and used them to construct a probability distribution from which we drew the initial stellar mass, disc mass, and infall rate.
More details on this procedure are given in Sect.~3 of \citetalias{2021A&A...645A..43S}.
We also assumed that the infall rates are constant throughout the infall phase. This is a strong simplification, and we discuss it further in Sect.~\ref{sec:Discussion}.

\subsection{Infall locations}\label{sec:infloc}
A disc is formed because the angular momentum initially present in the MCC must be conserved.
Given some assumptions about the initial angular momentum distribution (e.g. a spherical MCC in uniform  rotation), the source term for the infall onto the disc can be obtained (semi-)analytically.
Such recipes have been applied in the past and there is a wealth of literature on them \citep[e.g.][]{1955ZA.....37..217E,1956MNRAS.116..351B,1977ApJ...214..488S,1984ApJ...286..529T,2010A&A...519A..28V}.
However, the star formation process is more complex than suggested by such simple models.
The disc sizes from the SPH simulation can, in principle, be used to determine where the infalling material needs to be added, as was done in \citetalias{2021A&A...645A..43S} (see Sect.~2.10.1 for details).
In particular, this implies that the location where the infalling material is added to the disc ($R_\mathrm{i}$) is increasing with time.
In \citetalias{2021A&A...645A..43S}, we find that this way of determining the infall location leads to discs that are much larger than observed Class~0 disc sizes \citep{2020ApJ...890..130T}.
In order to make this comparison, we compared the gas disc sizes at the end of the infall phase to the observed Class~0 dust disc sizes.
Due to their large sizes, almost half of these discs were found to fragment.
In \citetalias{2021A&A...645A..43S}, we also investigated cases with much smaller infall radii, as suggested by some non-ideal magneto-hydrodynamic (MHD) simulations \citep{Hennebelle2016}.
In this case, we found that all discs remain stable against fragmentation, and that the discs are much smaller than observed.
In \citetalias{2023A&A...669A..31S}, we identified the initial conditions that lead to early disc sizes compatible with observations.
We iteratively determined infall locations that lead to early disc sizes compatible with the Class~0 disc sizes from \citet{2020ApJ...890..130T} and found the efficiency of accretion heating that simultaneously reproduces the observed luminosities.
We used infall locations constructed in this way in the present work.

The infall location is of the order of \SI{10}{au}, with a dependency on the stellar mass, so a different value is used in each of the 100 stellar mass bins (see Sect.~\ref{sec:imf}).
The values used in the baseline population are given in Appendix~\ref{app:rinf}.

\subsection{Initial orbits}
When a clump is introduced into the disc (Sect.~\ref{sec:model}), its initial orbital elements must be defined.
This should be done in a way that reflects the stochastic nature of fragmentation \citep{2012MNRAS.421.3286P,Young2015,Xu2025}.
We assigned an initial separation at a random location in the region of the disc where $Q_\mathrm{Toomre} < 2$.
Since clumps are formed in spiral arms \citep[e.g.][]{2004ApJ...609.1045M}, they are expected to be on initially eccentric orbits.
We therefore assigned them a random eccentricity between \num{0} and \num{0.3} (strictly speaking, we assigned an angular momentum deficit, as explained in Sect.~2.3 of \citetalias{Schib2025}).
Clumps are expected to have nearly flat orbits \citep[e.g.][]{Kubli2025}, and we therefore assigned them a very small random inclination from \num{0} to \SI{0.03}{\degree}.
The value of the inclination is not very important, as long as it is not zero (as this would confine all orbits to a plane).
Clumps start at apoapsis, with a random argument of periapsis and longitude of the ascending node.

\section{Results for the baseline population DIPSY-0}\label{sec:res}

In this section, we present and discuss the main results of our study, which are the fundamental statistical properties of the baseline population DIPSY-0.
In this population, we used $M_\mathrm{F}$ as the initial fragment mass (Sect.~\ref{sec:model}).
The gas accretion rate onto companions is limited to \SI{e-3}{\mj\per\year} (see Sect.~\ref{sec:helio}).
A random number (from one to three) of clumps are inserted.
Early disc sizes are set to be consistent with observed Class~0 discs (\citealt{2020ApJ...890..130T},\citetalias{2023A&A...669A..31S}), and the values for the background viscosity and the rate of external photoevaporation are according to Table~\ref{tab:param}.
We discuss several variant populations, where we studied deviations from these assumptions, in Sect.~\ref{sec:add}. 

The present section is organised as follows:
In Sect.~\ref{sec:fate}, we discuss the possible fates of fragments.
In Sect.~\ref{sec:am}, we show the mass-semi-major axis diagram of surviving companions, in Sect.~\ref{sec:pmf} the companion mass function (CMF), and in Sect.~\ref{sec:sma} the distribution of semi-major axes.
Section~\ref{sec:ecc} focuses on the eccentricities and inclinations of the companions, Sect.~\ref{sec:mult} on their multiplicity.
In Sect.~\ref{sec:mstar}, we discuss the dependency of our results on the host stellar mass.
Section~\ref{sec:coll} finally looks at collisions and close encounters.

\subsection{Fates of fragments}\label{sec:fate}
Over the \num{100000} systems of the baseline population DIPSY\nobreakdash-~0, a total of $\approx$~\num{430000} fragments form.
Only a small fraction of the fragments (7400, \SI{1.7}{\%}) survive for \SI{100}{\kilo\year}.

\begin{figure}
  \includegraphics[width=\linewidth]{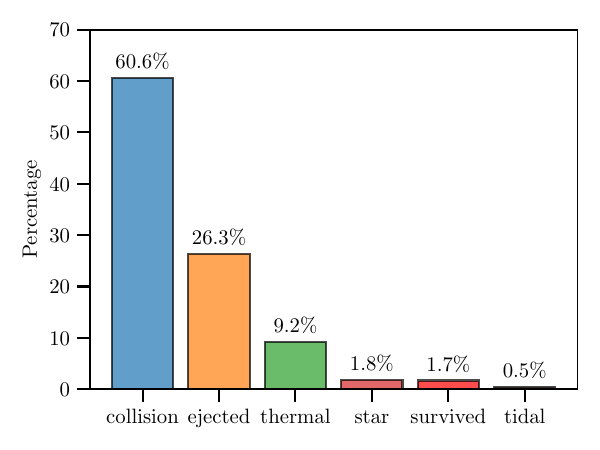}
  \caption{Occurrence of the different possible fragment fates for all fragments formed in the baseline population.
  From left to right: Collision and merger with another clump, ejection from the system, thermal disruption, accretion on the primary, survival of the fragment, tidal disruption.}
  \label{fig:bar_fate}
\end{figure}

Figure~\ref{fig:bar_fate} gives an overview of the possible fragment fates.
We find that the vast majority of objects (more than \SI{60}{\%}) are lost in collisions (see Sect.~\ref{sec:coll} for a detailed discussion). 
Another large group of objects (\SI{26}{\%}) is found to be ejected.
In this study, ejected objects correspond to those that move beyond \SI{10000}{au} from their host star. 
This typically occurs via gravitational interaction with other objects.
Another \SI{2}{\%} of objects are found to merge with the central star.
This occurs either because they migrate all the way across the inner edge of the grid or because they collide with the star via a highly elliptical orbit.
The fraction of clumps that are thermally disrupted is found to be \SI{9}{\%}.
Only a very small fraction of clumps (\SI{0.5}{\%}) are tidally disrupted.

\subsection{Mass-semi-major axis diagram}\label{sec:am}
\begin{figure*}
  \includegraphics[width=\linewidth]{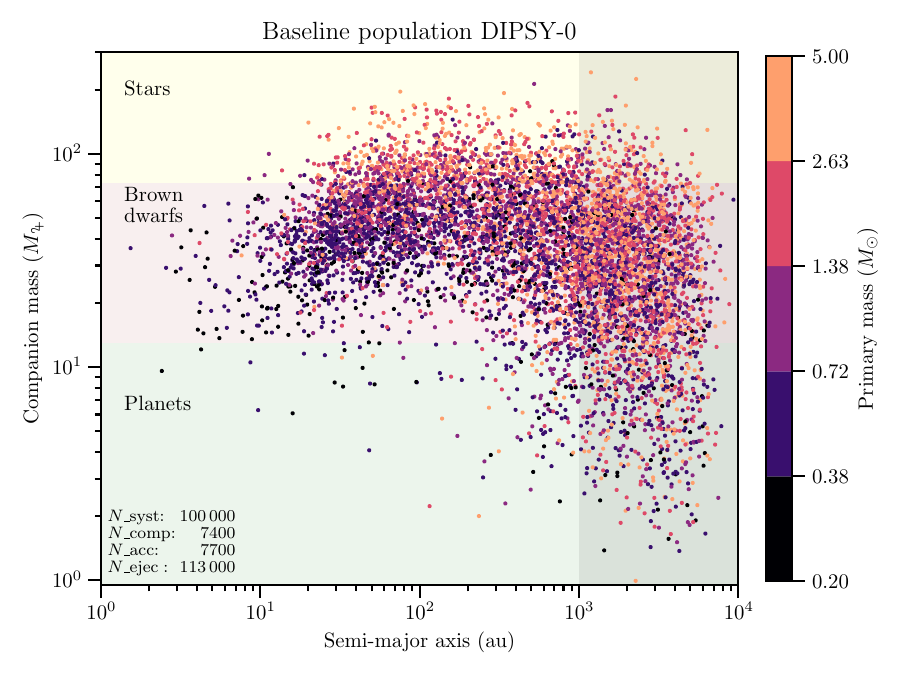}
  \caption{Mass vs semi-major axis for all the surviving companions of the baseline population.
  The colour code of each companion gives the (final) mass of the primary (host star).
  Differences in occurrence rate due to the different weights of stellar masses according to the IMF are not reflected in this figure; instead, for each of the 100 stellar mass bins, $\approx$~\num{1000} systems containing a total of 7400 companions are shown.
  The number of systems ($N_{\rm syst}$), the number of surviving companions ($N_{\rm comp}$), the number of objects accreted on the primary ($N_{\rm acc}$), and the number of ejected objects ($N_{\rm ejec}$) are given in the bottom left corner. 
  The horizontal background colours indicate how the companions might be classified as stars, BDs, or planets according to their mass. Clearly, the distinction between these classes is not as clear-cut as these limits may suggest.
  Objects at very large separations ($\gtrapprox\SI{1000}{au})$ may not survive long enough to be observed, as the grey area suggests, as they can become unbound.}
  \label{fig:am}
\end{figure*}

The \num{100000} systems were divided into \num{100} logarithmically spaced bins in final primary mass, ranging from \SIrange{0.05}{5}{\msun}.
This is necessary due to the strong mass dependence of the initial mass function (IMF) of stars, as we discuss in Sect.~\ref{sec:imf}.
A key result of population syntheses is the prediction of the mass-distance diagram, which might be of similar importance as the stellar HR-diagram \citep{2014prpl.conf..691B}. Figure~\ref{fig:am} shows the masses of all companions in the baseline population DIPSY\nobreakdash-0 against their semi-major axis. The semi-major axis is calculated in Jacobi coordinates.
The final mass of the primary (the host star) is indicated by the colour code in logarithmically spaced bins.
For these \num{100000} host stars, we find that \num{7400} companions survive.
They cover a large region in the mass-semi-major axis space, spanning five orders of magnitude in semi-major axis and two and a half orders of magnitude in mass.
Figure~\ref{fig:am} shows all the surviving companions of the baseline population.
The distribution of the underlying primary masses is approximately flat in $\log{M_\mathrm{*.final}}$, so does not follow the IMF.
In general, we find that the properties of surviving companions depend on primary mass.
Due to the binning, this dependence is not directly visible in Fig.~\ref{fig:am}.
We discuss the host stellar mass dependence in Sect.~\ref{sec:mstar}.

\subsubsection{Properties of the mass-semi-major axis distribution}
Overall, the envelope of the points covers a region that has roughly the shape of an inverted and rotated ‘L’ with a remarkable paucity of lower mass (planetary-mass) companions inside of 100 to 1000 AU. Within the region covered, there are no very strong pile-ups or sharp features. In orbital distance, the distribution is, roughly speaking, uniform in log. The mass has a more complex distribution that depends on orbital distance. We discuss the origin of the ‘L’ shape, which is a key result of this study, in Sec. \ref{sect:originofl}.

Although we find that all the companions share the same formation mechanism, it is still interesting to classify them according to their final mass into planetary-mass, BD, and stellar companions, because this can be observationally constrained. For this, we used the same mass boundaries between the planetary, BD, and/or stellar regime as \citet{Delorme2024}.

For such a classification, our results suggest that only about \SI{11}{\%} of all companions are planets (or planetary-mass companions), i.e. objects with masses $<$~\SI{13}{\mj}.
The vast majority (\SI{75}{\%}) of the objects are found to be BDs (\SIrange{13}{73.3}{\mj}), in agreement with previous studies (e.g. \citealt{Kratter2010,2012ApJ...746..110Z}).
Finally, we find that about \SI{14}{\%} of the fragments end up as stars.
The three classes are indicated through the different background colours in Fig.~\ref{fig:am}.
The least massive companion is a \SI{1}{\mj} giant planet, the most massive one a \SI{225}{\mj} (\SI{0.21}{\msun}) M-star.
The companion closest to the host star has a semi-major axis of \SI{0.15}{au}.
The furthest bound companions would be beyond our cutoff of \SI{e4}{au}.
We note that companions at very large separations (beyond $\sim$~\SI{1000}{au} as indicated by the grey shaded region in Fig.~\ref{fig:am}) might become unbound by stellar flybys or galactic tides \citep{Nordlander2017,PortegiesZwart2018,Bancelin2019,PortegiesZwart2021,PortegiesZwart2021a,Gratton2025}.
This effect is not easy to quantify in our simulations where each star-disc system forms and evolves in the absence of any dynamical interactions with the surrounding environment. The outcome will clearly depend on the age, surrounding stellar environment, and mass of the system. We encourage future studies to investigate this topic.

\subsubsection{The initial mass function of the host stars}\label{sec:imf}
In this study, we simulated the formation of companions via DI for a wide range of host star masses ranging from 0.05 to 5 $\msun$. This leads to special requirements when presenting the results, as stars of different masses have different occurrence rates.
It has long been known that stars with masses $>$~\SI{1}{\msun} are much less abundant than their lower mass counterparts (e.g. \citealt{1955ApJ...121..161S}).
The observationally determined IMF of the host stars is used to describe this dependency.
To have statistical significance (i.e. a high enough number of systems) also at higher host star masses, we divided the parameter space of final host star mass into \num{100} logarithmic bins in mass, and constructed our simulations such that each bin contains roughly \num{1000} systems.
When presenting the combined results obtained this way, we implicitly present them as if each stellar mass were equally likely in log, by discussing the population as a whole. 
For example, the companions shown in Fig.~\ref{fig:am} only show a preference for the primary mass to the degree that the fragmentation fraction has such a preference, \textit{not} because certain primary masses are more or less abundant according to the IMF.
The advantage is, however, that we can also see the outcome for massive stars, which are of particular interest but are much less frequent.

In the following, we discuss how the host star IMF influences our results, i.e. when the host stars are weighted according to the IMF.
We used the IMF given by \citet{2005ASSL..327...41C}:
\begin{equation}\label{eq:imf}
  \frac{\mathrm{d}n}{\mathrm{d} \log m} =
  \begin{cases}
  0.093 \times \exp \left\{ - \frac{1}{2}\frac{(\log m - \log 0.2)^2}{0.55^2} \right\}, & m \le \SI{1}{\msun}\\
  0.041 m^{-1.35}, & m \ge \SI{1}{\msun},
  \end{cases}
\end{equation}
In Eq.~\ref{eq:imf}, $\mathrm{d} n / \mathrm{d} \log m \equiv \xi(\log(m))$ denotes the stellar number density in $\si{pc^{-3}}$ per logarithmic interval of mass and $m$ is the stellar mass in \SI{}{\msun}.
\begin{figure}
  \includegraphics[width=\linewidth]{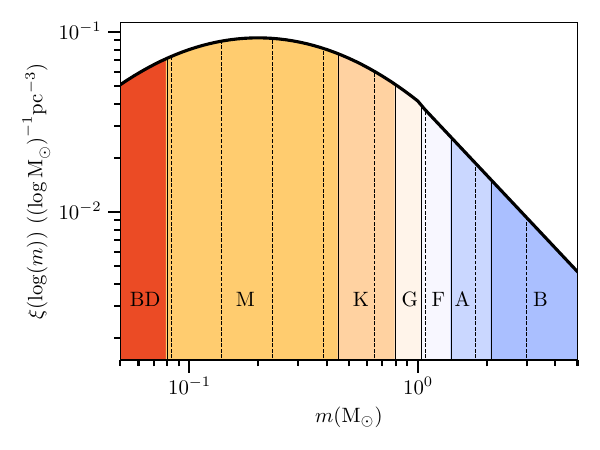}
  \caption{Initial mass function \citep{2005ASSL..327...41C} for the range of host star masses studied.
  The objects at the lower mass end are BDs, and the stellar spectral types from M to B are coloured\protect\footnotemark.
  Bins in final host star mass used in our study are shown as vertical dashed lines (only every tenth bin is shown for visibility).
  This demonstrates the range in primary mass covered in our study as well as the logarithmic binning.}
  \label{fig:imf}
\end{figure}
For unresolved systems, the IMF $\xi_\mathrm{sys}$ is slightly different below \SI{1}{\msun}:
\begin{equation}\label{eq:imfsys}
  \frac{\mathrm{d}n_\mathrm{sys}}{\mathrm{d} \log m} =
  0.076 \times \exp \left\{ - \frac{1}{2}\frac{(\log m - \log 0.25)^2}{0.55^2} \right\}, m \le \SI{1}{\msun}
\end{equation}
Instead of weighting each stellar mass bin equally as in Fig. \ref{fig:am}, we can also plot the mass-semi-major axis diagram inferred from our simulations for a region in space with a distribution of stellar masses according to the IMF, by weighting the systems with $\xi(\log m)$.

\begin{figure}
  \includegraphics[width=\linewidth]{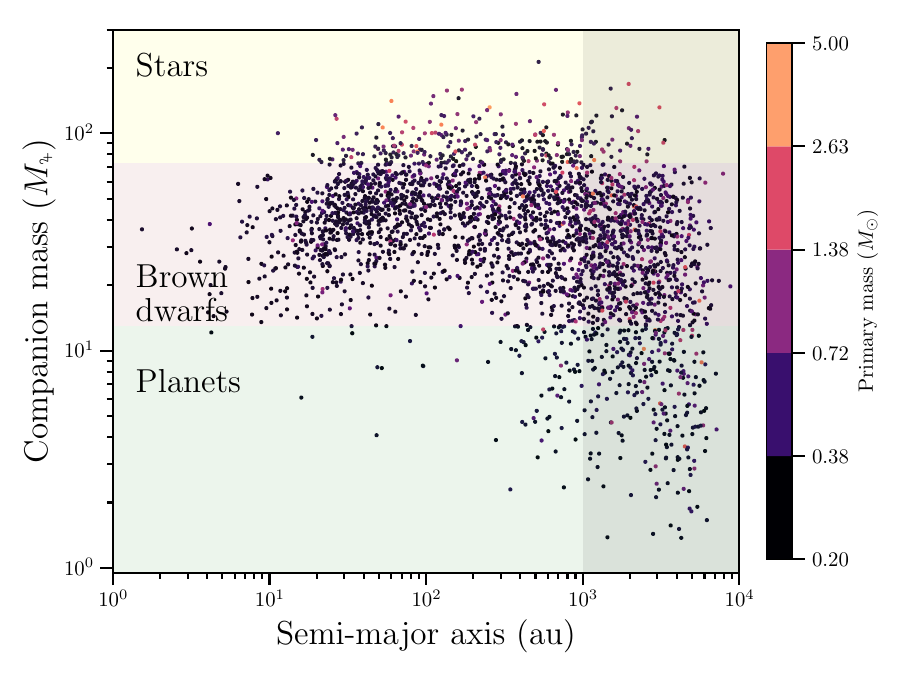}
  \caption{Mass-semi-major axis diagram, weighted according to $\xi(\log m)$ (Chabrier-2005 IMF, see text Sect.~\ref{sec:imf}).
  With this weighting, the population is completely dominated by M-dwarfs (note the difference in Fig.~\ref{fig:am}).}
  \label{fig:amimf}
\end{figure}

This is done by plotting a random selection of companions for each primary mass, where the fraction to be plotted depends on the stellar mass. This results in only a fraction of companions ($\approx$~\num{1000}) being shown, with almost all coming from the low mass bins. The result is shown in Fig.~\ref{fig:amimf}. We find that the mass-semi-major axis distribution is qualitatively similar to the one discussed in Sect.~\ref{sec:am}.
The strong preference for M-stars means the properties of this sub-population (discussed in Sect.~\ref{sec:mstar}) are dominant

The different weights need to be kept in mind when discussing statistical properties of the simulated population as a whole, except for properties with a weak primary mass dependence.
Using equal weights for each mass range results in averaged quantities that are not representative of observed distributions of stars.
On the other hand, quantities weighted as $\xi(\log m)$ are completely dominated by the systems of the lowest masses.
Therefore, in the following sections (Sect.~\ref{sec:pmf} to Sect.~\ref{sec:spectral}) we individually analyse narrower host star mass bins for the fundamental companion properties such as the mass or orbital distance distributions.

\subsection{The companion  mass function}\label{sec:pmf}
\begin{figure*}
  \centering
  \begin{subfigure}[t]{0.49\textwidth}
      \includegraphics[width=\linewidth]{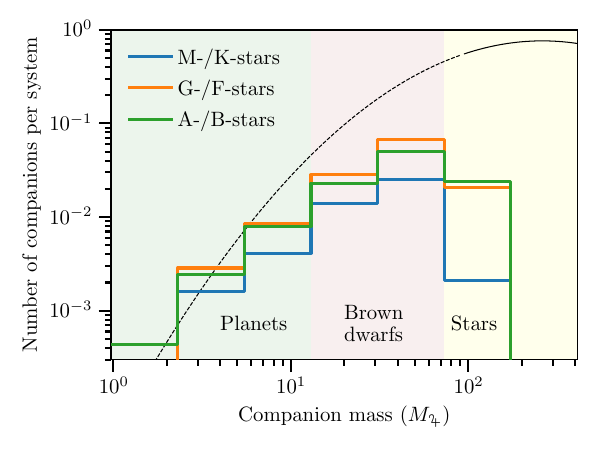}
  \end{subfigure}
  \hfill
  \begin{subfigure}[t]{0.49\textwidth}
      \includegraphics[width=\linewidth]{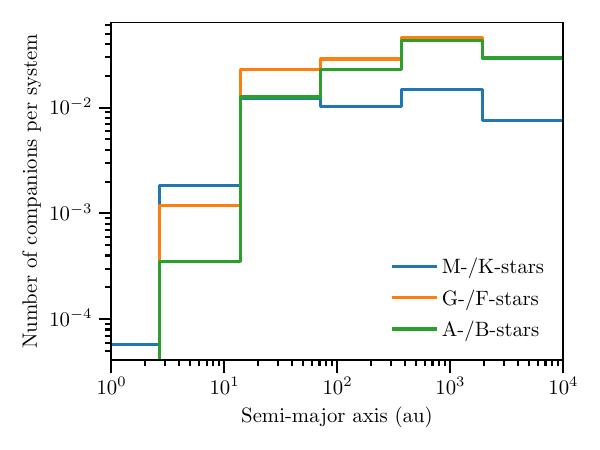}
  \end{subfigure}
  
  \vspace{2mm}
  
  \begin{subfigure}[t]{0.49\textwidth}
      \includegraphics[width=\linewidth]{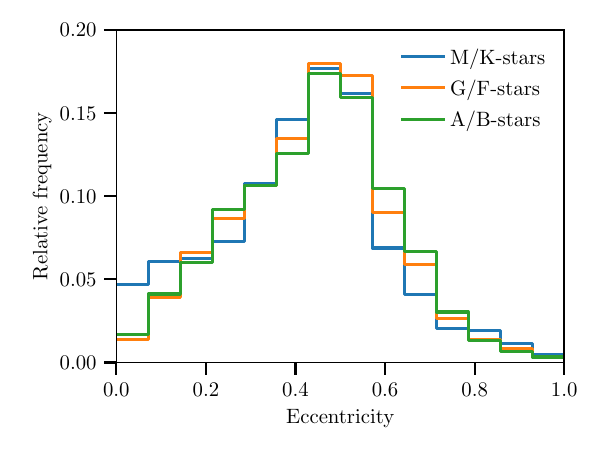}
  \end{subfigure}
  \hfill
  \begin{subfigure}[t]{0.49\textwidth}
      \includegraphics[width=\linewidth]{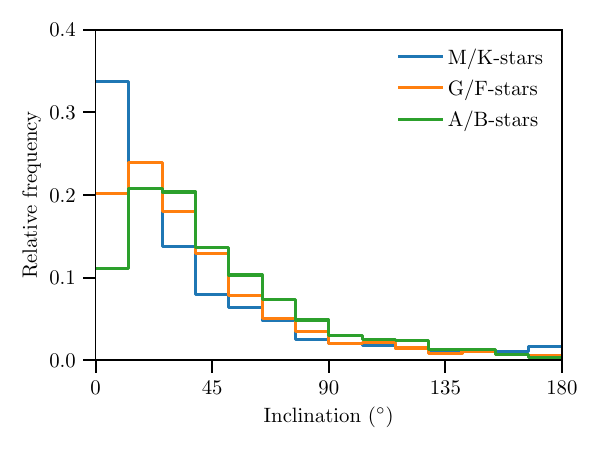}
  \end{subfigure}
  
  \caption{Distributions of four key properties for the companions of the baseline population DIPSY\nobreakdash-0, separately for M/K-, G/F-, and A/B host stars, weighted according to $\xi(\log m)$ within the three bins, for systems with $M_\mathrm{*,final} > 0.2 \msun$.
  Top left: CMF. For comparison, $\xi_\mathrm{sys}(\log m$) is shown as a thin black line (dashed part indicates extrapolation).
  Top right: Distribution of semi-major axes.
  Bottom left: Distribution of eccentricities. Bottom right: Orbital inclinations.}
  \label{fig:msei}
\end{figure*}
The top left panel of Fig.~\ref{fig:msei} shows the CMF for the baseline population. The distribution of the masses is a central quantity that carries important clues about the formation \citep[e.g.][]{Butler2006,Nielsen2010,2014MNRAS.439.2781M}.
The CMF is given separately for M/K-, G/F- and A/B-stars, respectively. Within the three bins, contributions from stellar hosts of different masses are weighted according to the host star IMF $\xi(\log m)$ as explained in Sect.~\ref{sec:imf}.
The shaded area indicates the type of companion according to mass, equivalent to Fig.~\ref{fig:am}.

\footnotetext{Credit: \\ \url{http://www.vendian.org/mncharity/dir3/starcolor/}.}

The CMF peaks at $\approx$\SIrange{30}{70}{\mj} in the massive BD range. The distribution is not symmetric, with a sharp drop-off towards the higher masses (above about 100-200 Jovian masses), and approximately a power law towards the smaller masses. 
In the range from \SIrange{2}{73.3}{\mj} we find for M/K\nobreakdash-stars: $N_\mathrm{M/K} \propto M_\mathrm{C}^{1.1}$ and for more massive stars: $N_\mathrm{G/F/A/B} \propto M_\mathrm{C}^{1.2}$.

The inferred CMF is found to be very similar across spectral types, indicating that dynamical processes (collisions and ejections) set it.
The only notable difference is at the high-mass end, where M- and K-stars have fewer companions.

Also shown in the figure is $\xi_\mathrm{sys}$, the IMF for unresolved systems (see Sect.~\ref{sec:imf}).
$\xi_\mathrm{sys}$ is only constrained down to $\approx$~\SI{90}{\mj}.
The extrapolation below this value is shown by the dashed line.
At lower masses, the CMF seems to follow $\xi_\mathrm{sys}$ approximately.
This result suggests that if DI is responsible for the formation of a fraction of the observed low-mass stars, then the BDs and planetary mass objects are just the low-mass tail of this star formation mechanism.
We note that our CMF may underestimate the number of stars formed by DI due to limitations in our model, which assumes that the companions formed always remain of lower mass than the primary (see Sect.~\ref{sec:nalim} and \ref{sec:helio}).

The CMF shows many objects in the brown dwarf desert (around \SI{30}{\mj}, e.g. \citealt{2006ApJ...640.1051G}).
However, the brown dwarf desert is observed at short orbital periods, while most companions in our population are on orbits $>> \SI{10}{au}$.
Our occurrence rates agree well with observations of distant companions (Sect.~\ref{sec:compo}).

\subsection{Distribution of semi-major axes}\label{sec:sma}
Similarly to the CMF, we can also investigate how the surviving companions are distributed in distance from the primary, which is also a fundamental quantity, as different formation mechanisms populate different regions \citep[e.g.][]{Fernandes2019,Fulton2021}.
The top right panel of Fig~\ref{fig:msei} displays the number of companions binned in semi-major axis, again in groups of spectral types. 

For G-/F-stars and A-/B-stars, the distribution is similar. Very few companions are found inside of 10-20 AU.
Outside, the distribution is approximately uniform in log, with a local maximum at several 100 to 1000 AU.
Companions around M- and K-stars seem to be more evenly distributed in log distances, with additional companions closer to the primary than for more massive host stars.
As mentioned, companions outside of roughly 1000 AU might become unbound due to gravitational interactions with other stars.

\subsection{Eccentricities and inclinations}\label{sec:ecc}
Next, we investigated two other important orbital elements \citep[e.g.][]{2020AJ....159...63B,Bryan2020,Bowler2023}, which are the orbital eccentricities $e$ and inclinations $i$ of the DIPSY\nobreakdash-0 population.
For multiple systems, a massive inner companion can cause the orbital elements of the outer companions to become meaningless in primary-centric coordinates.
Therefore, they are calculated in Jacobi coordinates.
The bottom panels of Fig.~\ref{fig:msei} show histograms of the eccentricities (left) and inclinations (right), respectively, of all companions.

As for the masses and distances, the eccentricities and inclinations shown in Fig.~\ref{fig:msei} are given as separate histograms for groups of primary spectral types: M- and K-stars, G- and F-stars, as well as A- and B-stars. Within the three bins the eccentricities and inclinations are weighted according to the IMF ($\xi(\log m)$) as described in Sect.~\ref{sec:imf}, and only systems with a final stellar mass $M_\mathrm{*,final} > 0.2$ are shown. Below this value, very few systems have surviving companions.

We find that the eccentricities are substantial, with typical values of \num{0.4}-\num{0.6}. They are found to be very similar across the large stellar mass range, with mean values of \num{0.41} to \num{0.44} and a standard deviation of $\approx$~\num{0.2}. We find additional companions with $0.0 < e \lesssim 0.2$ around M- and K-stars.
These high eccentricities are caused by the many scattering events occurring while many objects are in the system.

Inclinations are found to range between \num{20} and \SI{60}{\degree}, 
with mean values rising slightly with stellar mass from \num{35} to \SI{45}{\degree}.
There is a small number of companions with $i$ up to \SI{180}{\degree}. This is another indication that the final population is dominated by strong gravitational interactions between the (initially much more numerous) companions. 
Again, there is little difference across spectral types except for a tendency of more low-inclination orbits in M- and K-stars.
This is because in some lower mass systems, only one or a few fragments form, leading to less gravitational interaction.
Mean eccentricities and inclinations are given in Table~\ref{tab:ei}.
\begin{table}
\caption[]{Eccentricities and inclinations for groups of different spectral types.}
\centering
\begin{tabular}{ccccccc}  
\hline\hline
\begin{tabular}[c]{@{}c@{}}Spectral\\types\end{tabular} &
\begin{tabular}[c]{@{}c@{}}$e$\end{tabular} &
\begin{tabular}[c]{@{}c@{}}$i$ (\unit{\degree})\end{tabular} \\
\hline
M/K & \num{0.41+-0.19} & \num{36+-39} \\
G/F & \num{0.44+-0.18} & \num{39+-34} \\
A/B & \num{0.44+-0.18}   & \num{47+-34}   \\
\hline
\end{tabular}
\label{tab:ei}
\end{table}

\subsection{Multiplicity of the companions}\label{sec:mult}
\begin{figure}
  \includegraphics[width=\linewidth]{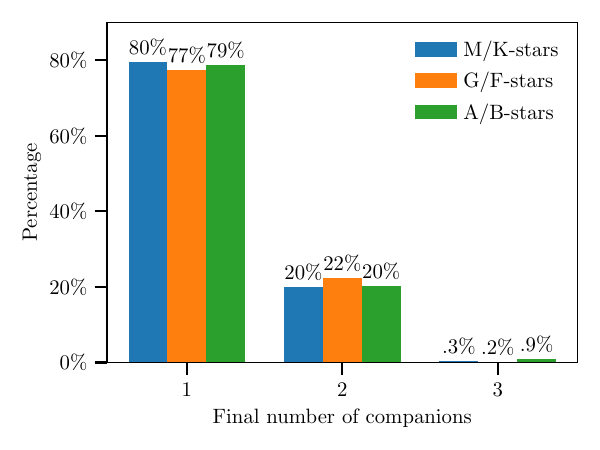}
  \caption{Companion multiplicity for different spectral types (for systems with at least one surviving companion).}
  \label{fig:mult}
\end{figure}

An important observational constraint for any formation model is the multiplicity of the resulting systems.
Here, we only consider systems with at least one surviving companion.
The maximum number of surviving companions is three, and Fig.~\ref{fig:mult} shows the fractions of systems with one, two, and three companions, again separately for M/K-, G/F-, and A/B-stars.
As seen in the figure, about 80\% of systems have only a single companion, while about 20\% have two. Thus, one surviving companion is the most likely outcome. Three surviving companions in a system are very rare (fraction of a percent).
Furthermore, the results are again similar for all the spectral types. We discuss the physical processes driving these results further in Sect. \ref{sect:nbinitialclumps}.

\subsection{Mass semi-major axis diagram as a function of primary spectral type}\label{sec:spectral}
\begin{figure*}[htb!]
  \centering
  \begin{subfigure}[t]{0.49\textwidth}
      \includegraphics[width=\linewidth]{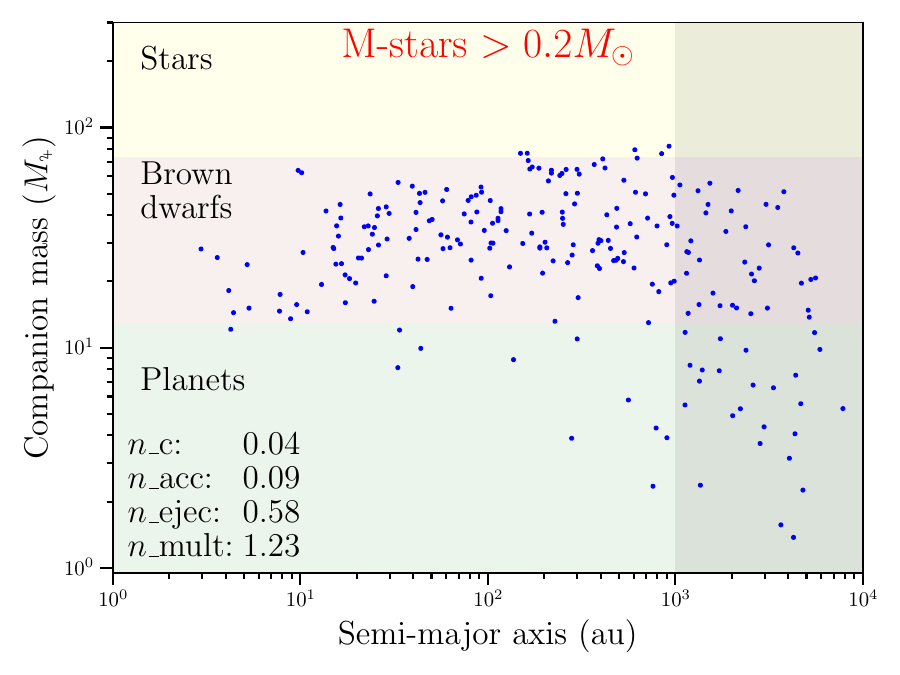}
  \end{subfigure}
  \hfill
  \begin{subfigure}[t]{0.49\textwidth}
      \includegraphics[width=\linewidth]{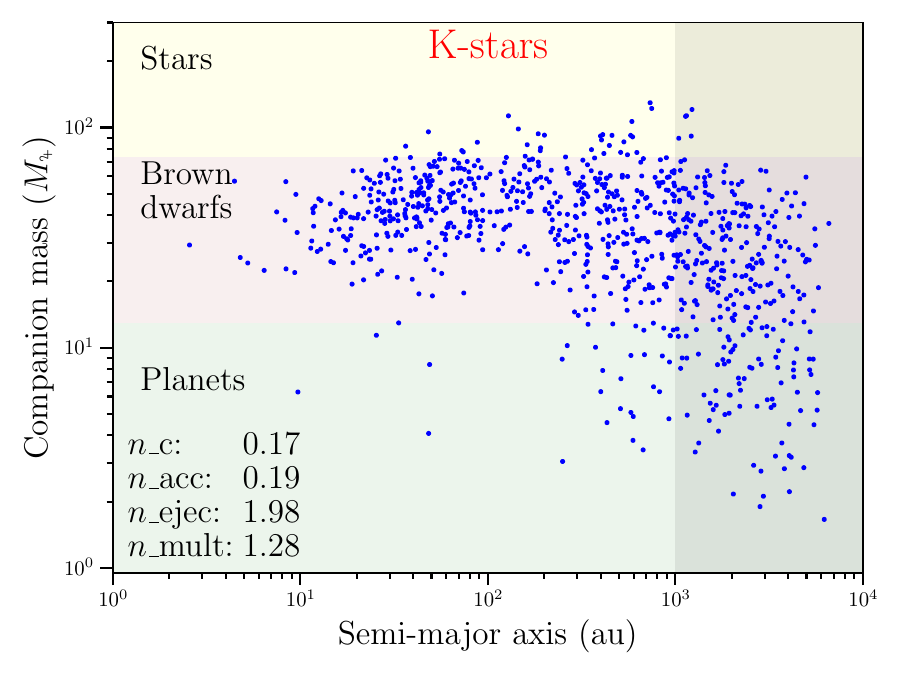}
  \end{subfigure}
  
  \vspace{2mm}
  
  \begin{subfigure}[t]{0.49\textwidth}
      \includegraphics[width=\linewidth]{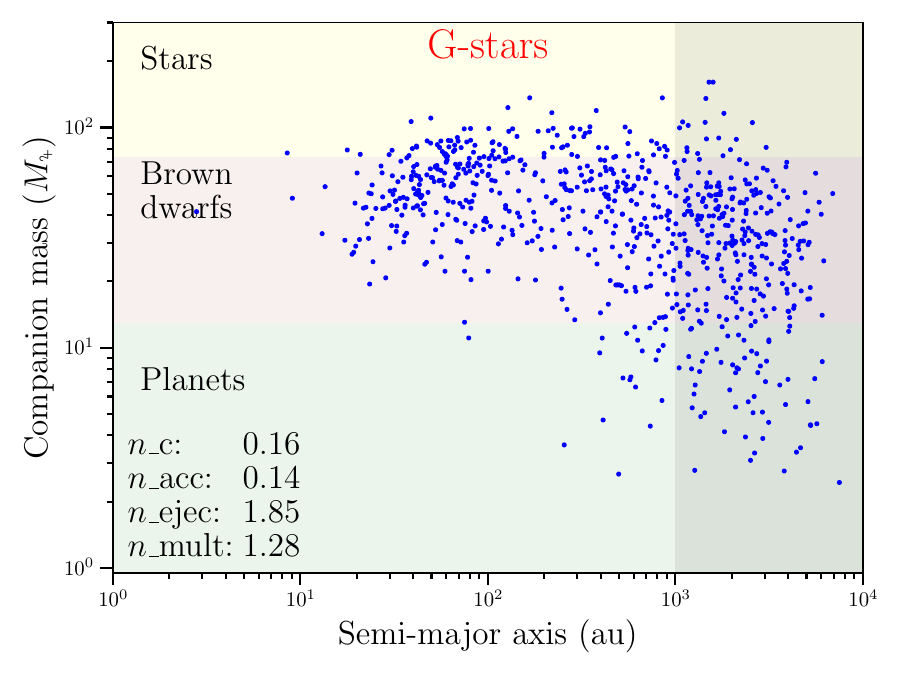}
  \end{subfigure}
  \hfill
  \begin{subfigure}[t]{0.49\textwidth}
      \includegraphics[width=\linewidth]{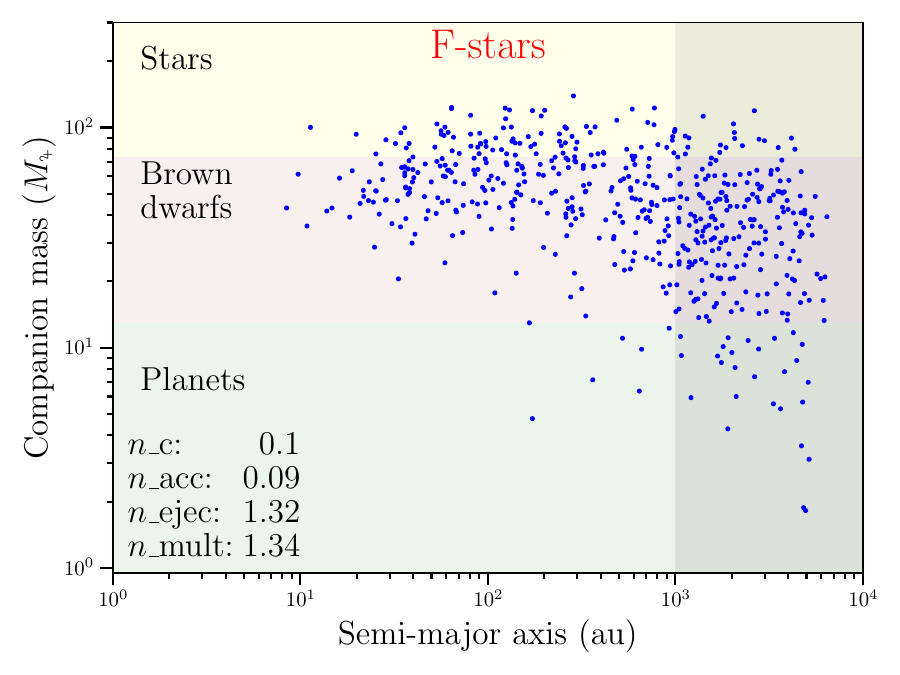}
  \end{subfigure}
  
  \vspace{2mm}
  
  \begin{subfigure}[t]{0.49\textwidth}
      \includegraphics[width=\linewidth]{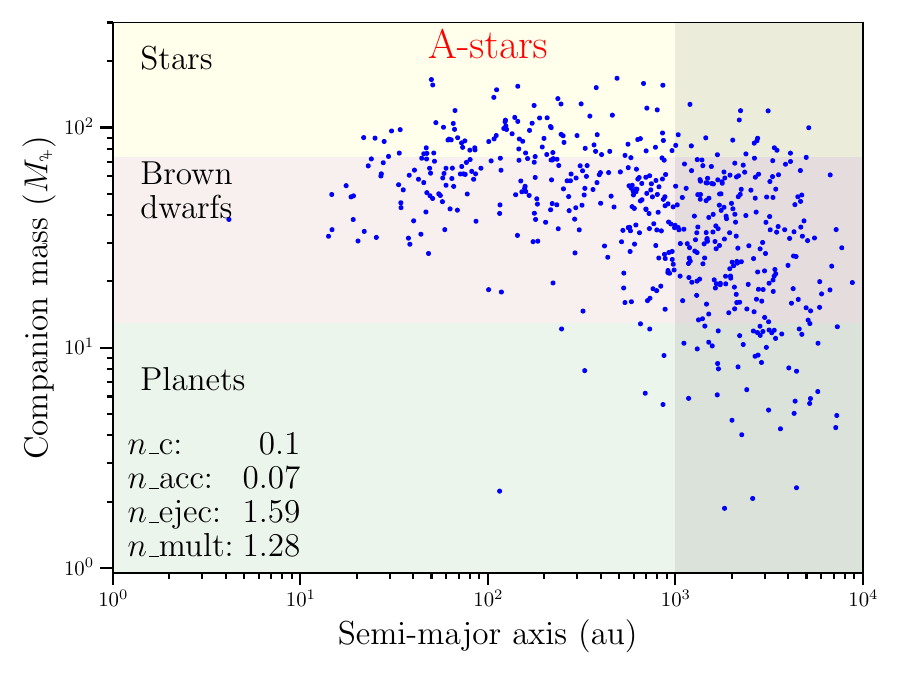}
  \end{subfigure}
  \hfill
  \begin{subfigure}[t]{0.49\textwidth}
      \includegraphics[width=\linewidth]{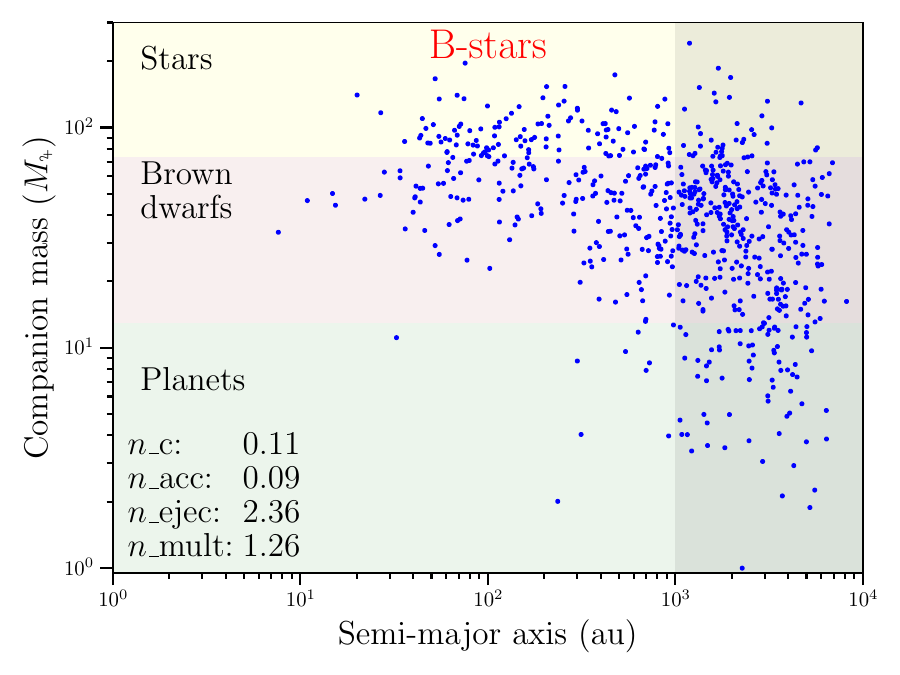}
  \end{subfigure}

  \caption{Mass vs semi-major axis for surviving companions of the baseline DIPSY-0 population for different spectral types of the primary.
  The spectral type, from M to B, is indicated in the title of each panel. 
  Each panel contains the companions from a random selection of \num{5000} systems from the corresponding mass range.
  In the bottom left corner, the following quantities are given as mean numbers for every sub-population:
  surviving companions per system $n_{\rm c}$, objects accreted on the primary per system $n_{\rm acc}$, ejected objects per system $n_{\rm ejec}$, and number of surviving companions per system with at least one surviving companion, $n_{\rm cs}$ (companion multiplicity). The envelope of points has a similar shape for all spectral classes but systematically shifts in position and density.}
  \label{fig:amspec}
\end{figure*}
In Sect.~\ref{sec:am}, we discussed the mass semi-major axis diagram for the surviving companions of all host star spectral types.
Due to the mass dependence of the IMF, it is difficult to compare the inferred populations for different primary spectral types when plotting them all together.
Thus, here we individually discuss the resulting populations for the various spectral types.
Figure~\ref{fig:amspec} presents the inferred populations for six different spectral types from M to B stars. 
The spectral types differ in the widths of the corresponding mass bins (Fig.~\ref{fig:imf}).
We therefore selected \num{5000} systems from each spectral type (without scaling according to the IMF inside each type) to be plotted in Fig.~\ref{fig:amspec}.
The different numbers of companions shown in the different panels reflect the dependence of the fragmentation fraction on the primary mass.

First, one notes that the rotated and inverted ‘L’ of the envelope of points covered in the mass semi-major axis diagram exists for all spectral types in a similar fashion, but that its exact locus and density shifts as a function of host star type.  
The top left panel shows companions around M-stars (\SIrange{0.08}{0.45}{\msun}).
We only selected systems with $M_\mathrm{*,final} > 0.2 \msun$ because very few systems with lower stellar mass fragment.
In the selected mass range, the mean number of surviving companions per system, $n_{\rm c}$,
is \num{0.04}.
The number of companions per system accreted on the primary, $n_{\rm acc}$, is roughly twice as high, and the number of ejected companions per system, $n_{\rm ejec}$ is almost $\num{0.6}$.
Among systems with at least one surviving companion, the mean number of companions (i.e. the companion multiplicity), $n_{\rm mult}$, is \num{1.23}.
Most companions around M-stars are found to be in the BD mass range, some planets can also form, they end up mainly at large distances, but almost none of the companions are in the stellar regime.

For K-stars (\SIrange{0.45}{0.8}{\msun}), $n_\mathrm{c}$ is about four times larger (\num{0.17}), resulting in additional points in the top-right panel.
The general distribution of the companions in the mass semi-major axis diagram is similar to that for M-stars, with slightly more massive companions at larger radial distances. 
$n_{\rm acc}$ is comparable to $n_{\rm c}$ here, $n_{\rm ejec}$ is larger than for M-stars ($\approx$~\num{2}).

The behaviour for the populations around G-, F-, A-, and B-stars is also found to be  similar.
There is some variation in the fragmentation fraction, as seen in $n_\mathrm{c}$, with a corresponding variation in $n_{
\rm acc}$ and $n_{\rm ejec}$.
The population shifts to more massive, more distant companions as the primary mass increases, leading to more stellar mass companions. 
We discuss this in Sect.~\ref{sec:mstar}.
A small fraction of planetary mass companions is inferred around primaries of all spectral types, especially at large separations.
We discuss this further in Sect.~\ref{sect:originofl}.
The quantities $n_\mathrm{c}$, $n_\mathrm{acc}$, $n_\mathrm{ejec}$ and $n_\mathrm{mult}$ for the various spectral types are summarised in Table~\ref{tab:n}.
\begin{table}
\caption[]{Companion properties for different spectral types.}
\centering
\begin{tabular}{ccccccc}  
\hline\hline
\begin{tabular}[c]{@{}c@{}}Spectral\\type\end{tabular} &
\begin{tabular}[c]{@{}c@{}}$n_{\rm c}$\end{tabular} &
\begin{tabular}[c]{@{}c@{}}$n_{\rm acc}$\end{tabular} &
\begin{tabular}[c]{@{}c@{}}$n_{\rm ejec}$\end{tabular} &
\begin{tabular}[c]{@{}c@{}}$n_{\rm mult}$\end{tabular} \\
\hline
M & \num{0.04} & \num{0.09} & \num{0.58} & \num{1.23} \\
K & \num{0.17} & \num{0.19} & \num{1.98} & \num{1.28} \\
G & \num{0.16} & \num{0.14} & \num{1.85} & \num{1.28} \\
F & \num{0.1} & \num{0.09} & \num{1.32} & \num{1.34} \\
A & \num{0.1} & \num{0.07} & \num{1.59} & \num{1.28} \\
B & \num{0.11} & \num{0.09} & \num{2.36} & \num{1.26} \\
\hline
\end{tabular}
\label{tab:n}
\end{table}

\subsection{Host star mass dependence}\label{sec:mstar}
So far, we discussed the properties of the baseline population as a whole or in rather large classes according to the spectral type.
Here, we systematically discuss the direct dependency of the inferred properties on the final stellar mass $M_\mathrm{*,final}$ and its physical reasons.
A larger $M_\mathrm{*,final}$ means a higher infall rate and/or a longer infall phase.
This typically leads to the formation of more fragments.
The larger mass reservoir also leads to more gas accretion.
Furthermore, discs around more massive stars tend to be larger \citepalias{2021A&A...645A..43S}, and fragments formed at larger distances are more massive (Eq.~35 in \citetalias{Schib2025}).
We can therefore expect that the inferred populations depend on $M_\mathrm{*,final}$, shifting to more massive and more distant companions with increasing mass.
Below, we discuss this dependency with respect to the masses, semi-major axes, eccentricities, and multiplicities of the companions. But first, we investigate how the prerequisites for companion formation (disc gravito-turbulence and fragmentation depend on host star mass).

\subsubsection{Gravito-turbulence, fragmentation, and survival}\label{sec:fractions}
Gravitational instabilities can play a role in systems even if they do not fragment.
In fact, almost all discs/systems in our baseline population have at least a short phase of gravito-turbulence ($Q_{\rm Toomre} < 2$) early in their evolution.
Only around the least massive primaries (less than 0.1 $M_\odot$), \SIrange{10}{20}{\%} of systems never reach this state.

The fraction of fragmenting discs is much lower.
Spiral arms driven by gravito-turbulence are very efficient in transporting angular momentum and mass, stabilising the discs against fragmentation \citep[e.g.][]{2004MNRAS.351..630L,Forgan2011a}.
The least massive systems do not fragment at all, only systems with $M_{\rm *,final} \gtrsim \SI{0.15}{\msun}$ will fragment.
The fragmentation fraction rises with primary mass up to \SI{0.6}{\msun} and then decreases again.
The increase is caused by the increasing disc mass around more massive primaries.
The decrease towards even more massive primaries is due to a combination of several effects.
An important factor is the early size of the disc.
However, it is unclear whether the observed sample of young discs we used to constrain the early disc sizes is indeed representative for stars with a final mass $\gtrsim~\SI{1}{\msun}$.
If massive stars are absent from the sample, we may under-predict the fragmentation fraction at high masses.
The ‘survival fraction’, i.e. the fraction of systems with at least one surviving companion, is again smaller, slightly more than half of the fragmentation fraction or between \SIrange{5}{15}{\%} with a weak dependency on $M_\mathrm{*,final}$.

\begin{figure}
   \includegraphics[width=\linewidth]{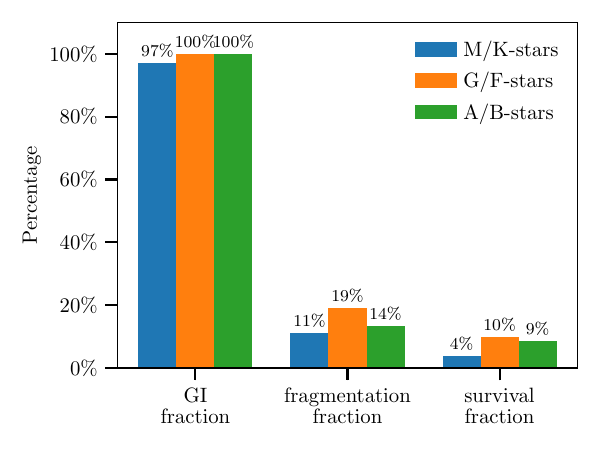}
\caption{Fractions of systems that are gravito-turbulent. These develop spirals (‘GI fraction’) that fragment and have at least one surviving companion.
The fractions are given for groups of spectral types according to the final host star mass $M_\mathrm{*,final}$.
All values of $M_\mathrm{*,final}$ \SIrange{0.05}{5}{\msun} are included.}
   \label{fig:fractions}
\end{figure}

In summary, above a final stellar mass of \SI{0.3}{\msun}, all the systems undergo a phase of GI, where \SIrange{10}{20}{\%} fragment, and about half as many have at least one surviving companion. In Fig.~\ref{fig:fractions} we visualise these results, showing these three quantities as fractions of the total number of systems in bins of spectral type.

\subsubsection{Companion mass}\label{sec:depmcomp}
\begin{figure*}
  \centering
  \begin{subfigure}[t]{0.49\textwidth}
      \includegraphics[width=\linewidth]{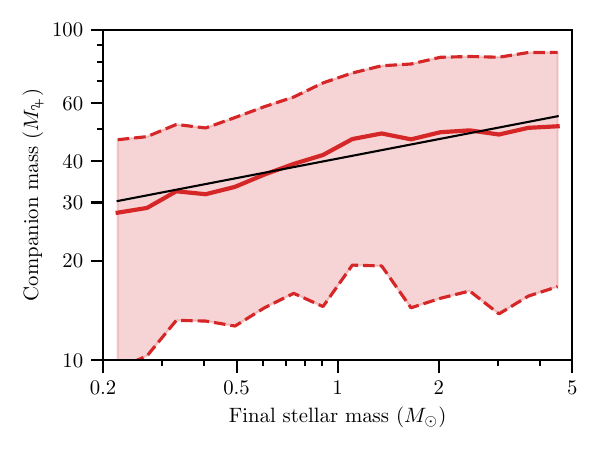}
  \end{subfigure}
  \hfill
  \begin{subfigure}[t]{0.49\textwidth}
      \includegraphics[width=\linewidth]{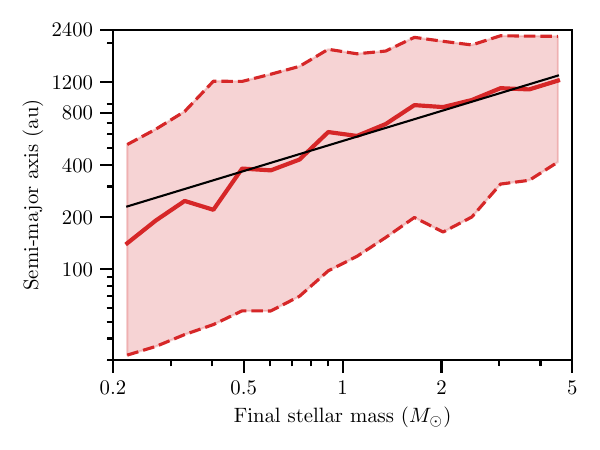}
  \end{subfigure}
  
  \vspace{2mm}
  
  \begin{subfigure}[t]{0.49\textwidth}
      \includegraphics[width=\linewidth]{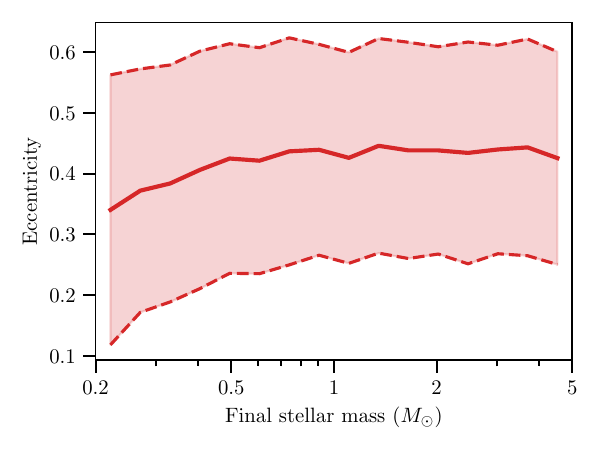}
  \end{subfigure}
  \hfill
  \begin{subfigure}[t]{0.49\textwidth}
      \includegraphics[width=\linewidth]{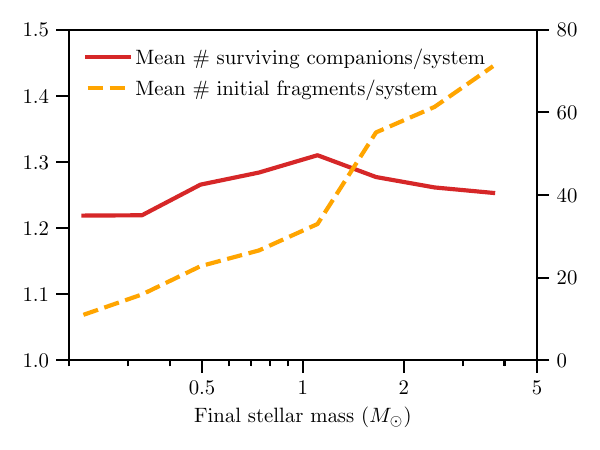}
  \end{subfigure}
  
  \caption{
  Quantities as a function of $M_\mathrm{*,final}$.
  Top left: Mean final companion mass, $M_\mathrm{c}$.
  The thin black line shows the best-fit power law.
  Top right: Median final semi-major axis, $S_\mathrm{c}$.
  The thin black line shows the best-fit power law. 
  Bottom left: Mean final eccentricity.
  Bottom right: Mean number of surviving fragments (left y-axis) per fragmenting system. The mean number of fragments formed initially (i.e. emerging in total during the disc's lifetime) is shown in orange (right y-axis), also per fragmenting system.}
  \label{fig:mstar}
\end{figure*}
The top left panel of Fig.~\ref{fig:mstar} displays the mean \textit{final} companion mass as a function of final stellar mass.
The spread (\SI{1}{\sigma}) in the fragment mass distribution is shown as shaded region.
The mean companion mass is increasing with primary mass.
Even around \SI{0.2}{\msun} hosts, the mean companion mass is in the BD regime ($\approx$~\SI{30}{\mj}).
It increases to \SI{50}{\mj} for \SI{5}{\msun} hosts. This means that there is a positive correlation of the mean companion mass with the primary mass although the correlation is rather weak (increase by a factor of about 1.7 for a 20-fold increase in the host star mass). The best-fit power law $~{M_\mathrm{c}/\mj\,=\,40\,(M_\mathrm{*,final}/\msun)^{0.2}}$ is shown in the top left panel of Fig.~\ref{fig:mstar}.
Interestingly, the distribution of masses becomes somewhat wider around more massive hosts ($\gtrsim~\SI{1}{\msun}$).
This happens because more massive systems tend to produce more fragments, and fragments formed later grow less in mass since the bulk of the available gas is accreted by their older, more massive counterparts. The effect is also seen in Fig.~\ref{fig:am}, where many of the lower mass companions are seen to orbit around the most massive stars.

\subsubsection{Semi-major axes}\label{sec:depsma}
In the top right panel of Fig.~\ref{fig:mstar}, the median semi-major axes $S_{\rm c}$ are shown vs.~final stellar mass, with the dashed lines indicating the first and third quartile.
The figure shows that, on average, companions are further from the primary with increasing primary mass.
The median semi-major axis is $\approx$\SI{150}{au} for the least massive primaries and increases to $\approx$~\SI{1200}{au} for \SI{5}{\msun} primaries.
The top right panel of Fig.~\ref{fig:mstar} also shows the best-fit power law $~{S_\mathrm{c}/\unit{au}\,=\,550\,(M_\mathrm{*,final}/\msun)^{0.6}}$. 

\subsubsection{Eccentricities}\label{sec:depecc}
It is also interesting to investigate the dependency between the inferred eccentricities and the primary mass.
The bottom left panel of Fig.~\ref{fig:mstar} depicts the mean companion eccentricities, including spread, after \SI{100}{Myr}.
The figure shows that there is only a very weak dependence, with the mean eccentricity growing from $\approx$~\num{0.34} to $\approx$~\num{0.44}.
This is caused by the increasing gravitational interaction between companions as the number of fragments increases with host mass.

\subsubsection{Multiplicities}\label{sec:depmult}
Since more mass is added to the disc during a longer infall phase in systems with a higher final stellar mass, such systems typically fragment more often. Fragmentation is also possible further out in the disc if the infall lasts longer and the disc can spread outwards while it is still self-gravitating.

In the bottom right panel of Fig~\ref{fig:mstar}, the dashed orange line (using the right y-axis) shows the mean initial number of fragments per fragmenting system (i.e. the mean number of clumps formed in total during the disc's lifetime). This demonstrates that there is an increase from about 10 to 70 fragments with stellar mass increasing from 0.25 to 3 $M_\odot$.

Given this trend, one wonders if more massive host stars also have additional surviving fragments in the final state. For higher primary masses, more fragments form, and also of higher mass. However, both of these aspects mean that there are also more gravitational interactions. Typically, in a system where strong gravitational interactions are important (like here because massive clumps emerge relatively close to each other), the system in the end only has a few (one or two, rarely three) companions, independently of the initial number, because of dynamical stability requirements \citep{Ford2001,Matsumura2010}. Interestingly, this mechanism is also observed in Class IV systems of dynamically active giant planets in a population synthesis model for CA \citep{Emsenhuber2023}.

This self-limiting effect is indeed visible in the bottom right panel of Fig.~\ref{fig:mstar} with the thick red line, which is the mean number of surviving companions per fragmenting system (left y-axis). It is essentially constant as a function of host star mass. The value is $\approx$~\num{1.2} around \SI{0.2}{\msun} stars and $\approx$~\num{1.3} around \SI{1}{\msun} stars. The slight decrease towards the most massive stars is probably caused by our treatment where companions beyond \SI{10000}{au} are considered as ejected objects. The comparison of the two curves also shows that only a very small fraction of all fragments survive.

This result on the final companion number (and the generally weak dependencies on host star mass) suggests that the outcome is not mainly driven by external processes and boundary conditions (like the stellar mass), but by physical processes happening inside of the system itself. As discussed further in Sect. \ref{sect:nbinitialclumps}, the insight that dynamical processes (collisions, scatterings, and ejections) are very important for the final outcome of DI is an important result. It is driven by the high number of fragments that form overall and then interact. This is difficult to fully capture for radiation-hydrodynamic simulations as they typically simulate shorter time spans (\citealt{2009ApJ...695L..53B};\citetalias{2018MNRAS.475.5618B};\citealt{Boss2021,Xu2025}).

\subsection{Collisions and close encounters}\label{sec:coll}
The gravitational interaction between clumps is modelled using the \texttt{mercury} N-body integrator (\citealt{Chambers1999}, Sect.~4.2 in \citetalias{Schib2025}).
We assume that all companions are point masses, as long as their closest approach exceeds the sum of their outer radii. This excludes the possibility of tidal interactions in close encounters. In addition, the point-mass assumption is clearly an over-simplification as clumps are very extended, have a lot of mass far from their centre, and are likely not spherically symmetric. Therefore, in reality, close encounters could strip some of the clumps' mass, or even unbind them (see Sect.~\ref{sec:disc_coll} for further details). Given that the radius of each companion is known at every point in time, \texttt{mercury}'s collision handling algorithm can be used to detect when two objects collide. We found that this happens very often (Sect.~\ref{sec:fate}). In the $\approx$~\num{12000} systems that fragment with at least two fragments formed, $\approx$~\num{260000} collisions are registered. Among these, $\approx$~\num{231000} (\SI{90}{\%}) happen between pairs of clumps, i.e. before either of the two has undergone a second collapse. This is expected, given that the radii of the clumps before the second collapse are very large ($\sim \unit{au}$), so that their size is not negligible compared to the size of their orbit.

All the collisions are assumed to lead to perfect merging. If two companions collide, they are combined into one object, which has the sum of the masses and momenta (inelastic collision). This is a strong assumption that might not always be appropriate (see further discussion in Sect.~\ref{sec:disc_coll}).

\subsubsection{Impact velocities}\label{sec:ivel}
\begin{figure*}
  \centering
  \begin{subfigure}[t]{0.49\textwidth}
      \includegraphics[width=\linewidth]{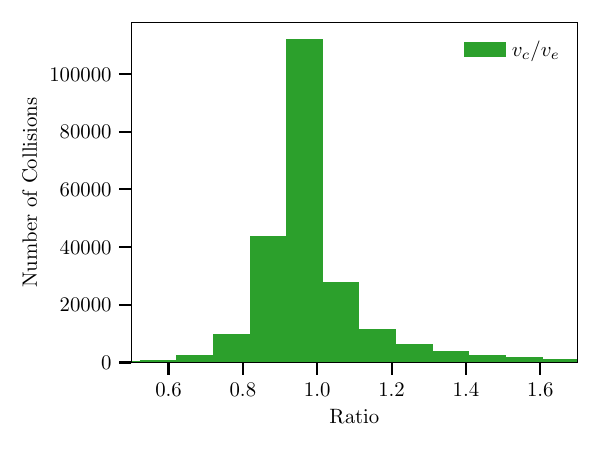}
  \end{subfigure}
  \hfill
  \begin{subfigure}[t]{0.49\textwidth}
      \includegraphics[width=\linewidth]{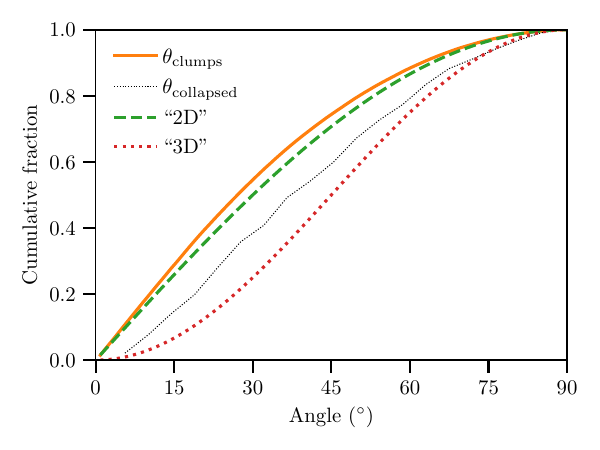}
  \end{subfigure}
  \caption{Left: Distribution of the ratio, $v_\mathrm{c}/v_\mathrm{e}$, for all clump-clump (non-collapsed objects only) collisions (see text Sect.~\ref{sec:ivel}) 
  Right: Distribution of the impact angle, $\theta$, for all clump-clump collisions. The solid orange line shows only collisions between clumps, while the thin dotted black lines shows those between collapsed objects.
  The expected distribution for random orientations (dotted red line) as well as for a flat system (dashed green line) are also shown.
  Further explanations are given in Sect.~\ref{sec:iangle}.}
  \label{fig:coll}
\end{figure*}

We found that the majority of collisions occur at low velocity.
The left panel of Fig.~\ref{fig:coll} shows a histogram of the quantity $v_\mathrm{c}/v_\mathrm{e}$,, the ratio of relative velocity at impact to mutual escape velocity, for all clump-clump collisions.
It reveals that most collisions (\SI{70}{\%}) actually happen at or slightly below escape velocity.
Only \SI{4}{\%} have $v_\mathrm{c}/v_\mathrm{e} > 1.5$. 
This happens due to the large sizes of the clumps and the strong damping that occurs on timescales comparable to or shorter than orbital timescales.

\subsubsection{Impact angles}\label{sec:iangle}
Another important quantity is the impact angle, as it can greatly influence the outcome of a collision.
The right panel of Fig.~\ref{fig:coll} shows a cumulative histogram of impact angles for all clump-clump collisions.
As seen in the figure, more than 75\% of the collisions occur at impact angles less than \SI{45}{\degree}.
This makes the assumption of perfect merging more reasonable.
Also for collisions where at least one body has undergone a second collapse, the perfect merging assumption seems to be reasonable \citep[e.g.][]{2012ApJ...745...79L,2021MNRAS.502.1647C}.
In the figure, the expected distribution of impact angles for a system of objects with random orientations (‘3D’) and for a flat system (‘2D’) is also displayed \citep{1962BookShoemaker}.
The clumps are initially close to the disc's midplane (though with small random inclinations).
We could therefore expect the distribution of impact angles to lie between the ‘2D’ and ‘3D’ curves.
We note that the statistics based on $v_\mathrm{c}/v_\mathrm{e}$ and $\theta$ is usually used for compact objects on Keplerian orbits.
Both assumptions are not necessarily appropriate for clumps, and therefore, the results may not agree with what is expected from collisions in planetary systems.
We discuss collisions further in Sect.~\ref{sec:disc_coll}.

\section{Variant populations}\label{sec:add}
Several of the underlying physical processes occurring during DI have not yet been studied in great detail with dedicated models. Thus, to calculate the baseline population, we had to make several uncertain assumptions that could significantly affect our results.
In the following, we investigate the effect of some of these assumptions using variant populations.
We simulated a total of six additional variant populations as summarised in Table~\ref{tab:add}.
\begin{table}
\caption[]{Additional populations with various parameters.}
  \centering
  \begin{tabular}{lll}  
  \hline\hline
  \begin{tabular}[c]{@{}c@{}}name\end{tabular} &
  \begin{tabular}[c]{@{}c@{}}description\end{tabular} \\
  \hline
  DIPSY\nobreakdash-1 & no limit on gas accretion \\
  DIPSY\nobreakdash-2 & no gas accretion, higher initial fragment mass \\
  DIPSY\nobreakdash-3 & no initial eccentricity\\
  DIPSY\nobreakdash-4 & larger initial disc sizes\\
  DIPSY\nobreakdash-5 & only one clump per fragmentation event\\
  DIPSY\nobreakdash-6 & lower viscosity and stronger disc photoevaporation\\
  \hline
  \end{tabular}
  \label{tab:add}
  \end{table}
  
\begin{figure*}[htb!]
  \centering
  \begin{subfigure}[t]{0.49\textwidth}
      \includegraphics[width=\linewidth]{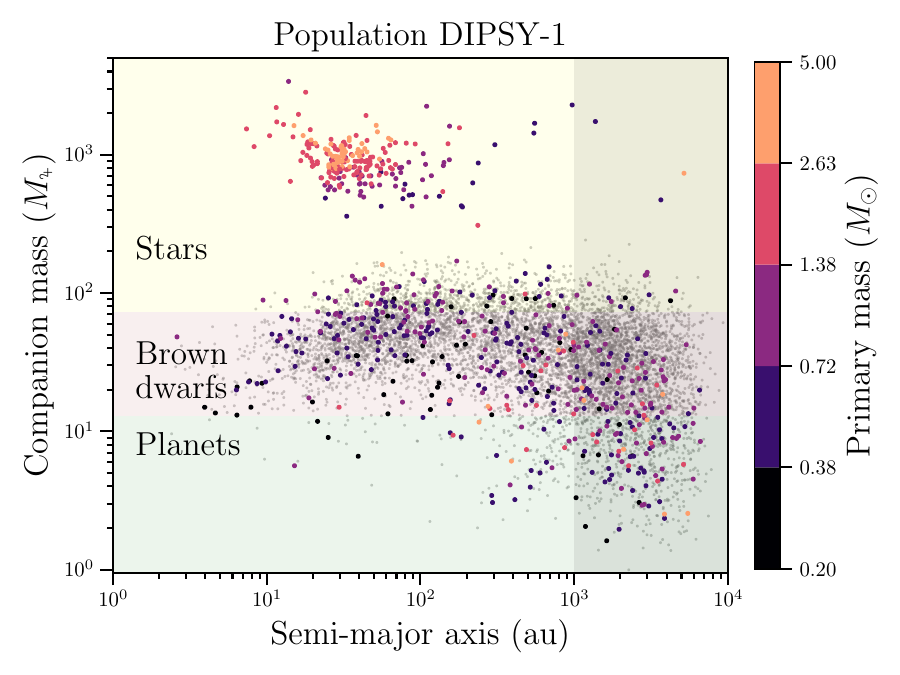}
  \end{subfigure}
  \hfill
  \begin{subfigure}[t]{0.49\textwidth}
      \includegraphics[width=\linewidth]{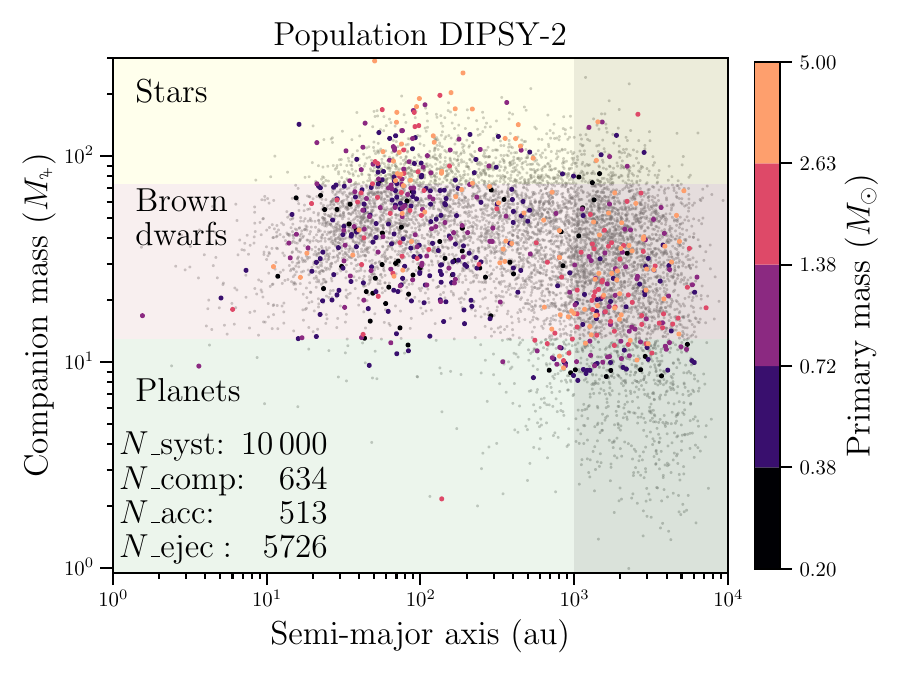}
  \end{subfigure}
  
  \vspace{2mm}
  
  \begin{subfigure}[t]{0.49\textwidth}
      \includegraphics[width=\linewidth]{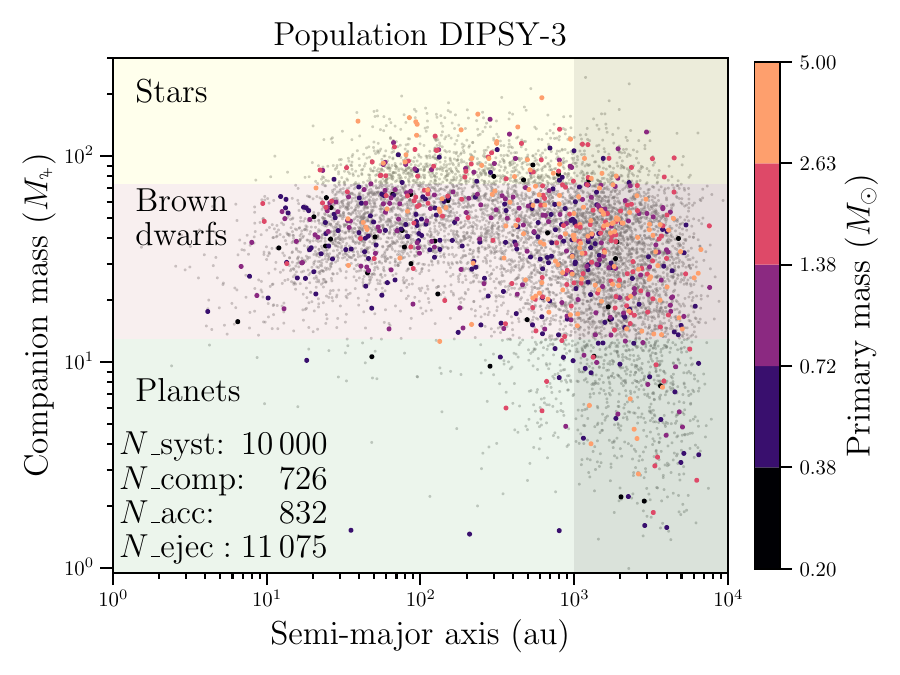}
  \end{subfigure}
  \hfill
  \begin{subfigure}[t]{0.49\textwidth}
      \includegraphics[width=\linewidth]{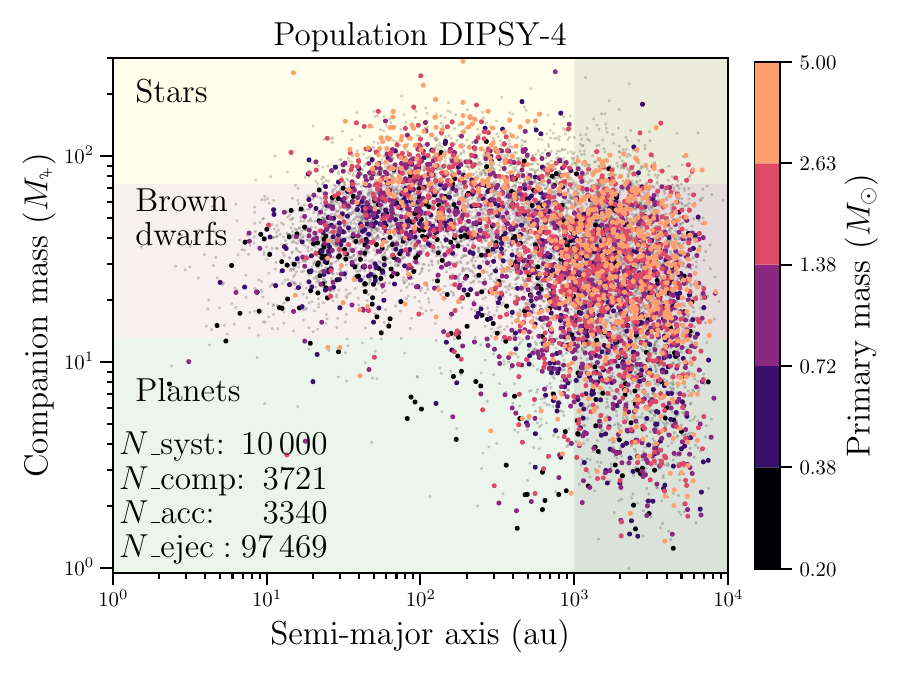}
  \end{subfigure}
  
  \vspace{2mm}
  
  \begin{subfigure}[t]{0.49\textwidth}
      \includegraphics[width=\linewidth]{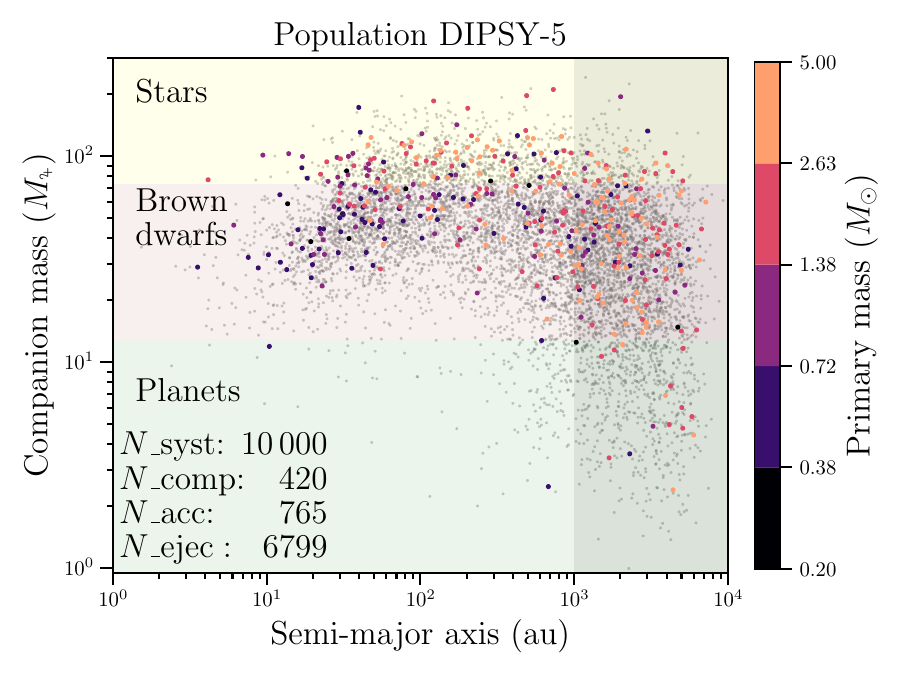}
  \end{subfigure}
  \hfill
  \begin{subfigure}[t]{0.49\textwidth}
      \includegraphics[width=\linewidth]{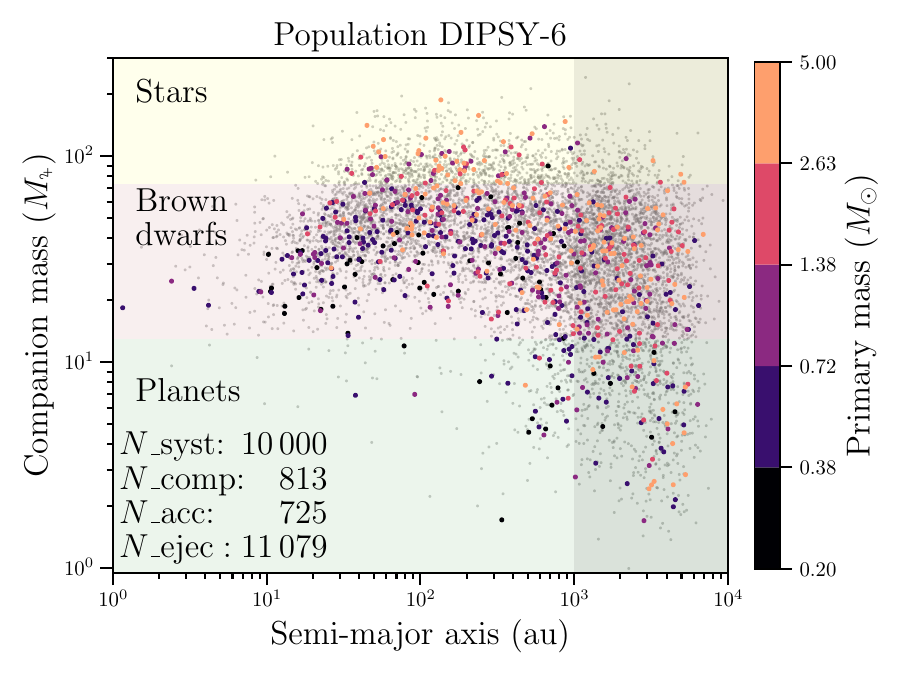}
  \end{subfigure}
  
  \caption{Mass vs semi-major axis (like Fig.~\ref{fig:am}) for the variant populations DIPSY\nobreakdash-1~to~6 as indicated in the panel titles.
  The baseline population DIPSY\nobreakdash-0 is shown in grey in the background.
  Please note that the top left panel shows the result of an experimental calculation in which some systems are unphysical because the secondary mass is large compared to the primary mass -- a situation not correctly captured by our model. Apart from this aspect, the variant populations are remarkably similar to the baseline one, indicating that internal processes dominate the outcome.}
  \label{fig:add}
\end{figure*}

Each of the populations listed in Table~\ref{tab:add} consists of \num{10000} systems, constructed as the baseline population, except for the specific changes we are going to discuss.
The mass-semi-major axis diagrams are shown in Fig.~\ref{fig:add}.

\subsection{DIPSY-1: No limit on gas accretion}\label{sec:nalim}
For the baseline population, we limited the rate of gas accretion onto companions to \SI{e-3}{\mj yr^{-1}}.
To study the effect of higher gas accretion rates, we removed the limit in population DIPSY\nobreakdash-1.
The results of this calculation need to be interpreted with caution.
Some of the simulations are not physical due to the underlying limitations of our model.
In assuming that the disc is always primary-centric, we implicitly assume that companion masses are much smaller than the primary.
With unlimited gas accretion, this is no longer valid.
In some cases, companions can become even more massive than the primary.
While the N-body interactions of said companion with the primary as well as other companions would still be correct, results related to migration, gas accretion, and the disc's gravitational potential will be unphysical.
The top left panel of Fig.~\ref{fig:add} shows the mass semi-major axis diagram for population DIPSY\nobreakdash-1, with the baseline population given as shaded dots in the background for comparison.

Without limiting the gas accretion rate, we see the emergence of a sub-population of stars with masses of \SIrange{0.5}{4}{\msun} at distances of 10 to 100 au that is absent in any of the other calculations.
Such stellar companions begin to dominate accretion when reaching a certain mass, and accretion on the primary is cut off.
However, as mentioned, our model cannot be used to realistically study such systems.
This experimental population is nevertheless helpful. It indicates that, depending on the precise nature of gas accretion, DI might form systems of binary or multiple stars of comparable masses.
It is recalled that for stars, multiple systems are very common, with a ratio of single:double:triple:quadruple of 1.50:1.0:0.105:0.026 \citep{Duquennoy1991}.

Interestingly, population DIPSY\nobreakdash-1 also shows a fairly high number of planetary mass companions.
It seems thus possible that this fraction would be higher in systems characterised by two or more stars of similar mass.

\subsection{DIPSY-2: No gas accretion and higher initial fragment mass}\label{sec:namj}
In this section, we explore the other limiting case of no gas accretion whatsoever. Gas accretion was not included in \citet{2018MNRAS.474.5036F} and \citet{2015MNRAS.454...64N}.
Additionally, we applied a higher initial mass for the fragments.
We used the local Jeans mass from \citet{2011MNRAS.417.1928F}:
\begin{equation}\label{eq:jeans}
  M_\mathrm{J,FR} = \frac{4}{3} \frac{2^{1/4} \pi^{11/4} c_\mathrm{s}^3 H^{1/2}}{G^{3/2} \Sigma^{1/2} \sqrt{1 + \beta^{-1}}}.
\end{equation}
In Eq.~\ref{eq:jeans}, $\beta = t_\mathrm{cool}\Omega$ is the local cooling time in units of the orbital timescale (Eq.~27 in \citetalias{2021A&A...645A..43S}).
$M_\mathrm{J,FR}$ is calculated in \citet{2011MNRAS.417.1928F} as a measure for the initial fragment mass in a self-gravitating disc and used as fragment mass in \citet{2018MNRAS.474.5036F}, see also Sect.~2.9.2 in \citetalias{2021A&A...645A..43S}.
In a Keplerian disc, it holds: $M_\mathrm{F} \approx 0.04 M_\mathrm{J,FR}$.
We use the higher initial mass here for comparability with \citet{2018MNRAS.474.5036F}, but also for computational reasons: no gas accretion means much more gas is available for later fragmentation and the number of fragments is increased a lot. 
The top right panel of Fig.~\ref{fig:add} shows mass versus semi-major axis of this population.

\begin{figure}
  \includegraphics[width=\linewidth]{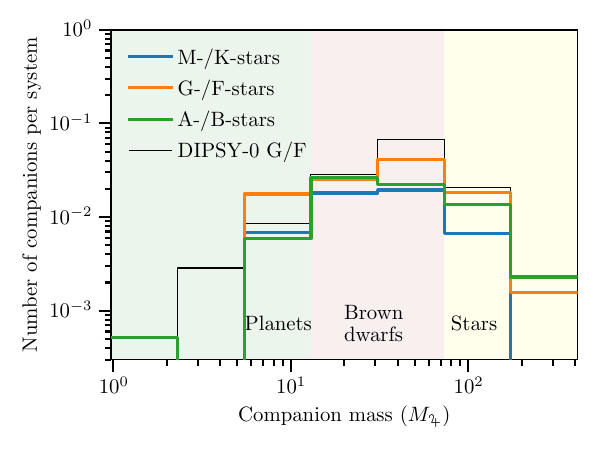}
  \caption{CMF for population DIPSY\nobreakdash-2 (similar to the top left panel of Fig.~\ref{fig:msei}), with no gas accretion and a higher initial fragment mass.
  The thin black line shows the baseline-CMF for G- and F-stars for comparison.}
  \label{fig:pmf_am2}
\end{figure}

The similarity of populations DIPSY\nobreakdash-2 and DIPSY\nobreakdash-0 is remarkable. While there is a stronger concentration at about \SI{100}{au}, the population without gas accretion overall occupies a similar mass semi-major axis space as the baseline population.
This is also seen in the CMF. Figure \ref{fig:pmf_am2} shows the CMF for population DIPSY\nobreakdash-2 
for G- and F-stars, the CMF from the baseline population is also shown for comparison.
The mass functions of the two populations are similar except for lower mass planets, which would not form due to the higher initial fragment mass.
The \SI{2}{\mj} companion at \SI{140}{au} is a rare case of a clump that survived tidal mass loss.
It appears that the higher initial fragment mass compensates for the mass gain through gas accretion.
The actual effect is more complex, though, since companions also grow through collisions in our work.

\subsection{DIPSY-3: No initial eccentricity}\label{sec:noecc}
In population DIPSY\nobreakdash-3, we studied the impact of assuming that clumps are formed on circular orbits instead of having a random eccentricity from \num{0} to {0.3} at birth, as in the baseline population.
The middle left panel of Fig.~\ref{fig:add} shows this population, again with the baseline population shaded in the back.

The figure does not reveal any clear difference from the baseline population.
The three $\approx1$\unit{\mj} companions inside of \SI{1000}{au} may indicate a tendency for additional closer-in low mass companions.
Compared to the baseline population, the fraction of surviving companions is slightly smaller, in favour of the fraction accreted onto the host star.
This may be because the migration rate tends to be higher on circular orbits.
We checked that the distribution of final eccentricities is also very similar to the baseline population, meaning that they are dominated by dynamical interactions rather than by the initial conditions. This is again a sign that the outcome is driven by internal processes among a highly dynamically and collisionally evolved population.

\subsection{DIPSY-4: Larger initial disc sizes}\label{sec:ldisc}
The next variant population, DIPSY\nobreakdash-4, has an infall location of the gas added onto the discs that is a factor of two larger compared to the baseline population.
This leads to much more fragmentation, because more mass is available in the gravitationally unstable region of the disc. 
The middle right panel of Fig.~\ref{fig:add} demonstrates this, showing the more than \num{3700} companions forming around 10,000 stars. For comparison, there were 7400 companions forming around 100'000 stars in the baseline population.

The mean number of surviving companions per system is thus five times higher than in the baseline population, demonstrating once again the importance of the disc size for fragmentation \citepalias{2023A&A...669A..31S}.
On the other hand, the distribution in mass semi-major axis space is very similar to DIPSY\nobreakdash-0.

\subsection{DIPSY-5: Only one clump per fragmentation event}\label{sec:onefrag}
In population DIPSY\nobreakdash-5, we only allowed the formation of one fragment in each fragmentation event instead of a random number from one to three in DIPSY\nobreakdash-0.
The envelope of points in the mass-distance diagram, shown in the bottom left panel of Fig.~\ref{fig:add}, is quite similar, but in terms of the number of fragments, only roughly half as many fragments per system survive.
This outcome is not fully obvious: while  only half as many fragments per fragmentation event form on average, the mass not removed will be available for future fragmentation.
The reason why fewer fragments survive is the reduced N-body interaction between them.
Having several fragments form at the same time makes their dynamical interaction more pronounced, and this tends to prevent some of these from undergoing rapid inward migration.
This again demonstrates the importance of N-body interactions.

\subsection{DIPSY-6: Lower viscosity and stronger disc photoevaporation}\label{sec:hia}
The next variant population, DIPSY\nobreakdash-6, is used to probe the influence of the background (minimal) $\alpha$-viscosity assumed in the simulations.
The baseline population uses a rather high value of the background $\alpha_\mathrm{BG} = \num{e-2}$. It is recalled that the actual (total) $\alpha$ is calculated based on the stability of the disc (see Sect.~2.2 of \citetalias{Schib2025}) and can be higher in the early phases.
At later times, this value might, however, be too high, and if accretion in protoplanetary discs is mainly driven by magnetohydrodynamic disc winds (see e.g. \citealt{2013ApJ...769...76B,2023A&A...674A.165W}), the remaining ‘background viscosity’ would be lower.
We test this by imposing $\alpha_\mathrm{BG} = \num{e-3}$ in population DIPSY\nobreakdash-6.
Additionally, we increased the rate of external photoevaporation by a factor of $\approx$\num{200} (see Sect.~2.7 of \citetalias{Schib2025}) to \SI{5e-6}{g.cm^{-1}.yr^{-1}} in order to reproduce the disc lifetimes seen in the baseline population.
The resulting population is shown in the bottom right panel of Fig.~\ref{fig:add}.

The resulting population is again  similar to DIPSY\nobreakdash-0, in terms of the mass semi-major axis distribution as well as in the number of surviving, accreted, and ejected companions per system.
This similarity may seem surprising, as such a different viscosity could influence the disc evolution significantly.
However, the early evolution of the fragmenting population is completely dominated by self-gravity.
In this phase, the viscosity is given by Eq.~8 in \citetalias{Schib2025}.
The background viscosity is only important for the longer-term evolution of the discs.

In summary, we find that none of the assumptions we tested in this section has a major impact on the resulting population.
The exception is the limit on gas accretion discussed in Sect.~\ref{sec:nalim}, which potentially restricts our study to the formation of systems with a primary that dominates in mass.
Of course, there could be other assumptions that would have a larger impact (Sect.~\ref{sec:Discussion}), and we hope to address this in future studies.

\section{Comparison to previous studies}\label{sec:comp}
As mentioned in Sect.~\ref{sec:intro}, various population synthesis projects in the DI paradigm have been presented. Below, we compare our results with previous studies.

\subsection{Forgan, Rice, and collaborators}\label{sec:cfr}
\citet{2013MNRAS.432.3168F} performed a population synthesis of self-gravitating disc fragmentation and tidal downsizing \citep{2010Icar..207..509B,2010MNRAS.408L..36N}.
Their model used pre-evolved discs around \SI{1}{\msun} stars and clumps are inserted at random locations in the outer disc.
This first study did not include N-body interactions, but this was added later in \citet{2018MNRAS.474.5036F}.
Their Fig.~2 shows a comparison of the inferred population with and without N-body interactions in terms of the distribution of final mass versus semi-major axis.
They found that the majority of the surviving objects have masses between \SIrange{1}{200}{\mj}.
The masses can at most be as large as the initial fragment mass $M_\mathrm{J,FR}$, since no gas accretion or merges were considered.

Without N-body interactions, most surviving companions have semi-major axes of \SIrange{10}{100}{au}, close to the birth locations but with possible inward migration.
The 2018-update demonstrates the importance of gravitational interaction, with an inferred population ranging from \SIrange{e-1}{e5}{au}.
The bulk of the population lies in a similar mass-semi-major axis space as our baseline population, i.e. their population is also dominated by massive giant planets and BDs at large semi-major axes. There is even a hint of a similar rotated 'L'- of the envelope of points in their mass semi-major axis diagram that is characteristic for our results, i.e. a presence of lower / planetary mass objects at distances outside of about 100 AU and an absence of such objects between 1 and 100 AU.  

The distribution of multiplicities they inferred (right panel in their Fig.~4) is similar to our result (Sect.~\ref{sec:mult}). They find a 5:1 ratio of single:double companions, while we obtain approximately a 4:1 ratio.

However, there are also significant differences: their simulations predicted the presence of a significant number of objects inside of 1 AU covering a large mass range, which is absent in our population. These companions are formed by inward scattering. Their mass function is bimodal with peaks at about 25 and 80 Jovian masses, whereas we obtain an unimodal distribution. The eccentricity distribution is also different: while ours peaks at 0.4 with a fall-off towards both lower and higher values, their distribution peaks at about 0.05 and is flat between 0.1 and 0.95.

\subsection{Nayakshin models}\label{sec:cn}
\citet{2015MNRAS.452.1654N} performed a population synthesis study of planet formation around solar mass stars in the DI paradigm including tidal downsizing.
Their surviving population mainly consists of two groups.
A group of rocky planets with masses between approximately \SIrange{0.1}{10}{\mearth} and separations from \SIrange{0.1}{10}{au}.
This group results from inward migration and tidal downsizing stripping the gaseous envelope of a solid core that forms via the accretion of solids by the gaseous clumps. Solid accretion is not included in our model, meaning that such a scenario cannot happen by construction (tidal downsizing in the sense of loss of gas only is included in our model, but not the formation of solid cores.).
The second group consists of giant planets (\SIrange{0.5}{10}{\mj}) that did not suffer such catastrophic mass loss of the gaseous envelope.
The separations of these objects lie predominantly between \SIrange{10}{90}{au}, with a fraction residing closer to the star (down to \SI{0.1}{au}).
This is a region of parameter space that is almost not populated in our study.
Their masses are also clearly lower (about 0.4 to 10 Jovian masses, with no objects more massive than 20 $\mj$), in strong contrast to our findings.
This is mainly due to the absence of gas accretion and mergers. 
Initial fragment masses are typically lower than in \citealt{2018MNRAS.474.5036F}).
Furthermore, while our companions often have large semi-major axes (typically at least tens of AUs or even hundreds of AUs), no objects are found in their simulations outside of 90 AU. Thus, there are significant differences between the predictions made by these two models.
The main reason for these differences is clearly that \citet{2015MNRAS.452.1654N} consider one fragment per system and consequently no N-body interactions.

\section{Comparison to observations}\label{sec:compo}
The majority of companions formed in our population are on wide orbits, with semi-major axes between \SIrange{e1}{e4}{au}.
Direct imaging surveys are therefore best suited to compare our results against.
A quantitative statistical comparison of our populations to observational surveys is not straightforward due to observational biases, the underlying assumptions such as the mass-luminosity relation, different sample properties, etc.
We thus perform an in-depth comparison with observations in a dedicated publication.
In the following, we compare our results only with a limited number of observational results, concentrating on the frequency of companions, a question which has been at the centre of attention in past surveys.  
In Sect.~\ref{sec:bn} we discuss the conclusions from a review of older surveys.
Sections \ref{sec:shine} and \ref{sec:beast} look at results from two specific, more recent surveys.

\subsection{Review by Bowler \& Nielsen 2018}\label{sec:bn}
A number of large imaging surveys have been conducted in the past. One of the first is \citet{Biller2013}. It is possible to perform meta-analyses of such surveys to constrain the occurrence rate of wide companions.
\citet{2018haex.bookE.155B} review the analyses available at the date of writing.
The authors conclude that the occurrence rate of massive giant planets (from \SIrange{5}{13}{\mj}) at separations between \SIrange{5}{500}{au} lies close to \SI{1}{\%}.
Our baseline population predicts a lower number in comparison. 
There is a population of planetary mass objects.
However, as is obvious from Fig.~\ref{fig:am}, most of these companions lie beyond \SI{500}{au}.
For the limits in mass and semi-major axis given above, we find companion fractions of \SIrange{0.4}{0.6}{\permil} around all primary spectral types, so approximately 20 times less.
\citet{2018haex.bookE.155B} also give an occurrence rate between \SIrange{1}{4}{\%} of BDs (with masses from \SIrange{13}{75}{\mj}) at the same separations.
Here, the agreement with our baseline population is very good: we find \SIrange{2}{3}{\%}. This is a non-trivial result given that the model was not optimised to reproduce any observational constraints.

The only population we synthesised that has more planetary mass companions inside \SI{500}{au} is population DIPSY\nobreakdash-4 with a larger early disc size and thus more fragmentation.
In this population, $\approx$~\SI{0.5}{\%} of systems have a companion with a mass of \SIrange{5}{13}{\mj} between \SIrange{5}{500}{au}.
However, the fraction of systems with a BD companion is also larger $\approx$(\SI{10}{\%}) than what is found observationally by \citet{2018haex.bookE.155B}, showing that in our model, we cannot simply adjust one quantity without (undesired) effects for others.
An overview of these numbers is given in Table~\ref{tab:BN}.
\begin{table}
\caption[]{Comparison to occurrence rates from a meta-analysis.}
  \centering
  \begin{tabular}{ccccccc}  
  \hline\hline
  \begin{tabular}[c]{@{}c@{}}Spectral\end{tabular} &
  \multicolumn{2}{c}{BN18} &
  \multicolumn{2}{c}{DIPSY-0} &
  \multicolumn{2}{c}{DIPSY-4} \\
  \begin{tabular}[c]{@{}c@{}}types\end{tabular} &
  \begin{tabular}[c]{@{}c@{}}Pl\end{tabular} &
  \begin{tabular}[c]{@{}c@{}}BD\end{tabular} &
  \begin{tabular}[c]{@{}c@{}}Pl\end{tabular} &
  \begin{tabular}[c]{@{}c@{}}BD\end{tabular} &
  \begin{tabular}[c]{@{}c@{}}Pl\end{tabular} &
  \begin{tabular}[c]{@{}c@{}}BD\end{tabular} \\
  \hline
  M & \multirow{3}{1.5em}{\SI{1}{\%}} & \multirow{3}{3em}{\nolinebreak{1 - 4 \%}} & \SI{0.6}{\permil} & \SI{2.3}{\%} & \SI{0.7}{\%} & \SI{10.5}{\%}\\
  K/G/F & & & \SI{0.6}{\permil} & \SI{2.9}{\%} & \SI{0.4}{\%} & \SI{7.2}{\%}\\
  A/B & & & \SI{0.4}{\permil}   & \SI{2.2}{\%} & \SI{0.4}{\%} & \SI{12.7}{\%}\\
  \hline
  \end{tabular}
  \tablefoot{Occurrence rates of distant giant planets (Pl) and BDs given in \citet{2018haex.bookE.155B} (BN18) as well as for two different DIPSY populations (see text Sect.~\ref{sec:bn}).}
  \label{tab:BN}
  \end{table}

\subsection{SHINE project}\label{sec:shine}
The SHINE (SPHERE Infrared Survey For Exoplanets) project is a large survey that uses the SPHERE instrument at the Very Large Telescope \citep{Desidera2021,Langlois2021}.
\citet{2021A&A...651A..72V} perform a statistical analysis of a subsample of 150 stars with spectral types from B to M.
The authors use the BEX-COND-hot evolutionary tracks \citep{Marleau2019} to infer the mass of the observed companions.
Using a parametric model, they predict the frequencies of systems with at least one companion with a mass between \SIrange{1}{75}{\mj} and a separation from \SIrange{5}{300}{au}.
Their parametric model also attributes a frequency to a ‘planet-like’ (i.e. CA) formation pathway called ‘PPL/LN’ and a ‘binary star-like’ formation pathway called ‘BDB’, which might also be fuelled by DI.
The results from \citet{2021A&A...651A..72V}, together with the results from our model in the same mass/semi-major axis limits, are summarised in Table~\ref{tab:shine}.
\begin{table}
\caption[]{Comparison to the SHINE survey.}
\centering
\begin{tabular}{ccccccc}  
\hline\hline
\begin{tabular}[c]{@{}c@{}}Spectral\\types\end{tabular} &
\begin{tabular}[c]{@{}c@{}}full\end{tabular} &
\begin{tabular}[c]{@{}c@{}}BDB\end{tabular} &
\begin{tabular}[c]{@{}c@{}}DIPSY \end{tabular} \\
\hline
M & \num{12.6}$^{+13}_{-7}$ & \num{5.4}$^{+8.7}_{-4.4}$ & \num{2.1} \\
K/G/F & \num{5.8}$^{+4.7}_{-2.8}$ & \num{3.2}$^{+3.0}_{-1.8}$ & \num{2.5} \\
A/B & {23.0}$^{+14}_{-10}$   & \num{4.1}$^{+4.2}_{-3.0}$ & \num{1.7}   \\
\hline
\end{tabular}
\tablefoot{Frequencies of systems with at least one companion from the SHINE survey \citep{2021A&A...651A..72V}, compared with our baseline DIPSY-0 population.
The term ‘full’ denotes a parametric model that includes planet- and binary star-like formation, while ‘BDB’ only includes the latter (see text Sect.~\ref{sec:shine}).
}
\label{tab:shine}
\end{table}
The result from the SHINE survey predicts companion frequencies from binary-like formation at the percent level.
This agrees qualitatively with the prediction from DIPSY.
There is quantitative agreement within a factor of \num{2.5}, which is a fair agreement given the many rough model approximations involved, such as the low dimensionality of the model.
The trend with stellar mass is at this point unclear. \citet{2021A&A...651A..72V} state that their results seem to show a local minimum in the frequency of substellar companions around FGK stars. They caution, however, that this result should not be overinterpreted because their sample only contains 20 M-stars, and that the analysis of the full SHINE sample should yield more robust results. Also, within their 1-sigma error bars, the observed BDB-frequencies are compatible with a BDB-frequency independent of spectral type.

At the present time, the preliminary SHINE results are compatible with an increase in companion frequency from K/G/F to A/B stars, while we find a slight opposite trend (decrease by a factor of 0.68).
A possible explanation lies in our initial conditions.
The sizes of young discs are quite uncertain, especially as a function of stellar mass. In our simulations, the infall location is chosen such that (\citepalias{2023A&A...669A..31S} the observed early disc sizes as found by \citet{2020ApJ...890..130T} are reproduced. 
It is not known if the observed sample contains A/B~stars.

\subsection{BEAST survey}\label{sec:beast}
The BEAST (B-star Exoplanet Abundance Study, \citealt{2021A&A...646A.164J}) survey is a direct imaging survey also using the SPHERE instrument, but targeting B-stars and focusing on the detection of exoplanets.
\citet{Delorme2024} study the frequency of giant exoplanets by statistically analysing the first half of the BEAST survey.
The sample consists of 34 young B-stars with a median mass of \SI{4.8}{\msun}.
In the selection of this sample, stellar companions between $\approx$~\SIrange{12}{720}{au} are excluded.
When comparing to our population, we therefore also exclude such systems.
The occurrence rate of substellar companions around B-stars is given as \num{11}$^{+7}_{-5}\%$ by \citet{Delorme2024}.
Using the aforementioned constraints, the corresponding number from DIPSY\nobreakdash-0 is $\approx$~\SI{7}{\%}.
Again, given the many underlying simplifications, this seems a fair agreement.
However, while $\approx$~\SI{60}{\%} of the companions from DIPSY are massive BDs (\SIrange{30}{73.3}{\mj}), the two detected companions that are part of the BEAST survey are both at most low-mass BDs:
b~Centauri~b \citep{2021Natur.600..231J} has a mass of \SI{11+-2}{\mj}, $\mu^2$~Sco~B \citep{Squicciarini2022} a mass of \SI{14+-1}{\mj}.
In terms of the companion-to-star mass ratio, the discrepancy becomes even bigger.
It therefore appears that b~Centauri~b and $\mu^2$~Sco~B are not typical outcomes of DI as represented in our current model \footnote{We note that the numbers reported in Sect.~6.4 of \citet{Delorme2024} were obtained with a preliminary calculation of our model and therefore differ (slightly) from the figures given here}. This will be studied further in future work.
Future comparisons will also include other quantities, such as the orbital distance, multiplicity, or the eccentricity distributions \citep{Wahhaj2013,2020AJ....159...63B}.

\section{Discussion and limitations}\label{sec:Discussion}
We performed a population synthesis with a model that includes a significant number of self-consistently linked physical effects compared to some previous studies. While doing so, we still had to make a number of assumptions and simplifications.
In this section, we discuss some of the physical processes driving our results but also how the most important assumptions and simplifications could affect our findings.

\subsection{High number of clumps formed and consequences: Scatterings and collisions as internal processes}\label{sect:nbinitialclumps}
Our simulations agree with previous radiation-hydrodynamic studies in the finding that mass loading onto the protoplanetary disc from infall can drive the disc into multiple episodes of fragmentation \citep{2009ApJ...695L..53B,2012ApJ...746..110Z,2020A&A...644A..41O}. In such hydrodynamic simulations, which simulate (in comparison to our infall duration) often only shorter time intervals (several $10^3$ to a few $10^4$ years), the formation of a significant number of fragments can be observed especially if the mass loading is high and the infall radius is large. For example, in one simulation of \citet{2009ApJ...695L..53B}, 11 fragments form, while in the simulation of \citet{2012ApJ...746..110Z} with the highest mass loading rate and infall radius, half a dozen fragments emerge rapidly. In our simulations, the durations of the infall phase are chosen to lead to a stellar mass distribution following the IMF. While typical infall durations are a few $10^4$ years, there is also a non-negligible fraction of longer durations of $10^5$ years \citepalias{2023A&A...669A..31S}. These longer timescales are especially prevalent for more massive stars, because in our setup, in order to form more massive stars, the infall duration needs to be longer on average. 

This explains the (at first sight) non-trivial result, that in the baseline population, among 100,000 stars, a huge number of about 430 000 fragments form in total.
Given that only a minority of systems fragment (Sect.~\ref{sec:fractions}), fragmenting systems will have tens of fragments formed over their lifetime on average.
In Fig.~\ref{fig:mstar}, the mean number of fragments is shown as a function of the final host star mass. One can see a clear increase from about 10 at 0.2 $M_\odot$ to 70 at 30 at 3 $M_\odot$. This reflects the longer infall duration, allowing more fragments to form in total. 

This very high number of initially (i.e. in total) formed clumps needs to be sharply distinguished from the final number of companions per fragmenting disc, which is only about 1.2 with nearly no dependency on host star mass. This implies that the formation process is very inefficient at turning initial fragments into bound companions. Instead, almost all fragments are either destroyed in collisions, or (with a factor of $\sim$2 lower probability) ejected. With these results at hand, a new vision of DI arises where the final outcome is a strongly collisionally and scattering-dominated process. These ‘internal’ processes explain key outcomes such as the number of final companions or the eccentricities, and the weak dependency of many results on host star mass.

\subsection{Origin of the 'L' shape in the mass-distance plane}\label{sect:originofl}
\begin{figure}
  \centering
      \includegraphics[width=1\linewidth]{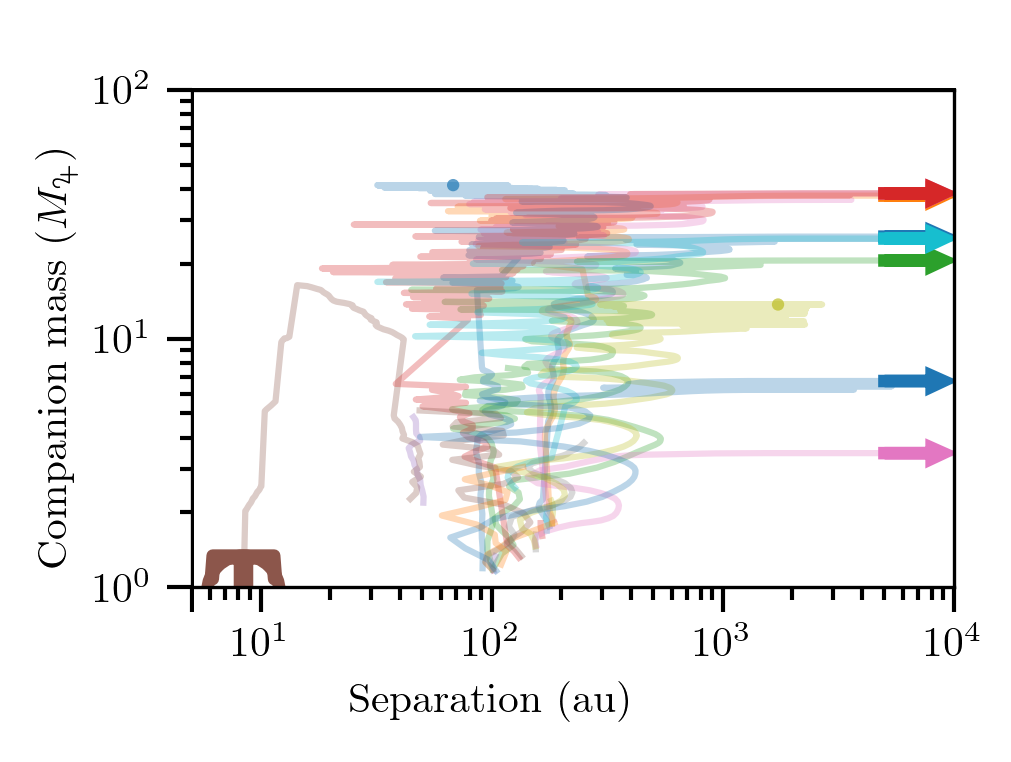}
  \caption{Example of the formation of a system with an inner more massive and an outer, less massive companion. The lines show the growth tracks of the clumps.
  Arrows to the right denote ejections, and the ‘T’ stands for the (ultimate) thermal destruction.}
  \label{fig:aMtracksexample}
\end{figure}

One of the key findings of this study is the prediction of the rotated and inverted ‘L’ envelope that is populated by the companions in the mass semi-major axis diagram with a paucity of planetary mass objects inside of \SIrange{10}{100}{au}. Figure \ref{fig:aMtracksexample} shows the formation tracks in an example system with two surviving companions of about 40 $\mj$ at 100 AU and a 13 $\mj$ at 2300 AU. The final primary mass is 1.06 $\msun$. Initially, 23 clumps emerge within the first 30 kyr between 50 and 200 AU with masses of \num{1} to \num{2} Jovian masses. They then grow by rapid gas accretion and collisions. Typically at $\sim$ $\mj$, the second collapse occurs. One clump forming inside of the others undergoes inward migration and is destroyed by the combined action of disc irradiation and tidal mass loss.
After 100 kyr, the system contains 9 collapsed clumps, which interact gravitationally. By 320 kyr, only the two final companions are left. From the growth track of the outer lower mass companion, one can see that outward scattering out of the gas-rich part of the disc leads to the formation of the lower mass object further out, while the clump scattered outside grows by gas accretion and, initially, collisions. Thus, in order to not grow to masses higher than planetary mass, the orbital distance needs to be larger than the discs' size of 100-1000 AU. Otherwise, they grow into the brown dwarf or even stellar regime. Thus, the combination of the emergence of many interacting clumps, scatterings, and gas accretion inside and the absence of it outside leads to the ‘L’ shape. 

We note that there is an interesting analogy in low-mass star formation: One hypothesis suggested that low-mass stars might form due to ejection from denser regions of a collapsing molecular cloud. While current understanding particularly emphasises that low-mass stars form from the turbulent compression and fragmentation in molecular clouds producing collapsing cores over a wide range of masses, it is still possible that ejection also plays a role depending on the formation environment (see e.g. \citealt{Luhman2012}).

\subsection{Infall model}
For the formation of the star-and-disc systems, we assumed a relatively short infall phase characterised by a high and constant infall rate at the beginning of the simulations (Sect.~\ref{sec:init}, Sect.~2.5 of \citetalias{Schib2025}).
While short episodes with high infall rates are also seen in hydrodynamic simulations (e.g. \citealt{2018MNRAS.475.5618B}), a number of observations show that accretion varies strongly during stellar mass assembly \citep{Fischer2023}.
Such accretion variability could modify our results.
For example, if the mass supply to the disc is dominated by episodes of high accretion rates that are separated in time, this could lead to fewer fragments being in the disc at the same time, which could reduce the number of surviving fragments.
At the same time, sudden drops in accretion rate will favour fragmentation since they drive the disc towards instability through the decreasing temperature.
The net effect of accretion variability on the outcome of DI is difficult to predict, since it depends on the precise nature of the variability, which also has an influence on the host's properties \citep{2018MNRAS.475.2642K}. We plan to investigate this topic further in the future.

\subsection{Collisions}\label{sec:disc_coll}
Collisions occur very often in our simulations and they play a key role in setting the final companion masses.
Clump-clump collisions are particularly frequent due to the very large sizes of the bodies (typically $\sim~\unit{au}$, which is orders of magnitude larger than ‘normal’ (collapsed) planets and BDs). In assuming perfect merging, we made a strong assumption for the outcome of collisions.
This base assumption is to some degree justified, as we observed low impact velocities around or even below the mutual escape velocity because of the very strong damping (see Sect.~\ref{sec:coll}).
This is a very particular collisional regime, different from what is typically seen in collisions between, for example, rocky planets.
Interestingly, a similar regime exists for mergers of asteroids that are also large (with respect to their semi-major axes) and damped by their debris disc \citep{Wimarsson2025}.
A very recent study \citep{Matzkevich2024} focusing specifically on collisions between clumps finds a number of different possible outcomes.
Clump-clump collisions can, for example, result in disruption, erosion, or hit-and-run, depending on their ages, masses, and impact angles.
In some cases, the collision can even trigger the dynamical collapse of a clump.
This outcome may be especially relevant for our study, since the collapse makes the clumps robust against tidal disruptions.
However, \citet{Matzkevich2024} find this phenomenon is most likely for evolved clumps, some of which would soon undergo dynamical collapse even without a collision.
Thus, we do not expect this outcome to change our results very much.
Furthermore, the collision outcomes for clump-clump collisions from \citet{Matzkevich2024} (their Figs.~9-11) suggest that in some cases perfect merging is a reasonable assumption for collisions below mutual escape velocity at intermediate impact parameter. Such collisions are found to be rather common in this work. Clearly, further investigations of this topic are required.  
It is difficult to judge in what way a more elaborate collision model would influence our results.
Less perfect mergers may decrease the clumps' growth, possibly keeping them at a lower final mass and making them more prone to tidal disruption.
However, regardless of the outcome of collision, the total mass involved needs to be conserved.
So if a lot of mass is returned to the disc after an erosive or disruptive collision, this gas would again be available for accretion or further fragmentation.

Another simplification related to the collision model are close encounters.
As long as the centres of two bodies do not approach closer than the sum of their radii given by the interior structure model, they interact gravitationally as point masses.
For clumps, this may not be a good approximation, since a large fraction of their mass is relatively far from their centre \citep{galvagnihayfield2013}.
and may be stripped/unbound in a close encounter.
To our knowledge, no hydrodynamic simulations exist at the moment that specifically study close encounters of clumps (though see Fig.~6 in \citealt{Xu2025}).
It will also be important to address this topic in the future.

\subsection{Primary-centric disc and gas accretion}\label{sec:helio}
The accretion of disc gas is among the major factors influencing the final population of companions.
At the same time, it is a complex, three-dimensional process, which is still not very well understood.
Hydrodynamical simulations of gas accretion onto giant planets do not agree very well on the accretion rates.
For example, \citet{2019MNRAS.486.4398F} perform a comparison of seven different hydrodynamical codes simulating the migration and gas accretion of clumps at wide separation.
They find a factor of several difference for the accretion rate of an initially \SI{2}{\mj} clump at \SI{120}{au} with identical initial conditions.
The differences are lower in the case of clumps with an initial mass of \SI{12}{\mj}, and accretion rates are of the order of \SI{e-3}{\mj.yr^{-1}}.
We used this value as an upper limit for the gas accretion rate.
This seems reasonable also when comparing to \citet{2013ApJ...767...63S}.
The authors perform radiation-hydrodynamic simulations to study the early evolution of circumplanetary discs (CPD) around companions formed through disc fragmentation.
They find that long-term accretion rates from the CPD to the companion reach \SI{3e-4}{\mj.yr^{-1}}.

Without the limit on the accretion rate, our accretion model would allow some companions to approach the mass of the primary (Sect.~\ref{sec:nalim}).
While this may be a pathway for the formation of binary- or multiple-star systems, it is problematic in the context of our model.
The assumption of a primary-centric disc only makes sense if the primary is much more massive than any of the companions.
If much higher accretion rates are possible (as seen e.g. in \citet{2012ApJ...746..110Z} at much higher infall rate or in \citet{2020A&A...644A..41O} in the case of massive star formation) this would lead to a shift of the barycentre, clearly invalidating the primary-centric disc assumption.
Our limit on the accretion rate therefore effectively restricts the probed parameter space to systems with a massive primary.

It is also possible that accretion onto clumps is inhibited due to effects not included in our model.
\citet{2013MNRAS.435.2099N} study the migration and accretion of clumps and use analytical and numerical arguments to show that radiative feedback should reduce the accretion of disc gas onto low-mass clumps.
A similar effect is seen in \citet{Stamatellos2015}.
\citet{2021NatAs...5..440D} investigate disc fragmentation in the presence of the spiral-driven dynamo \citep{2019MNRAS.482.3989R,2020ApJ...891..154D} and find that initial fragment masses are much lower than in the absence of the dynamo, and that gas accretion is stifled due to magnetic shielding in the self-gravitating phase.
The combination of lower (approximately an order of magnitude) fragment masses and suppressed gas accretion would likely influence our results.
However, it is not straightforward to say how, as any mass not bound in clumps would be available for further fragmentation or future accretion.

\subsection{Magnetic fields}\label{sec:magnet}
Magnetic fields do not only affect fragmentation and the accretion of gas, but also play an important role in the formation of the star-and-disc system.
We discuss this in detail in \citetalias{2021A&A...645A..43S}.
The effect of the magnetised collapse is indirectly taken into account in our model by setting the infall location in such a way that we retrieve early disc sizes in agreement with observations \citepalias{2023A&A...669A..31S}.
Clearly, a more physically motivated approach that quantitatively captures the physics of all non-ideal MHD effects and its impact on the infall location as well as the angular momentum would be desirable and opens a perspective for future work.

Additionally, magnetically driven disc winds are thought to play a key role in driving accretion and removing mass in protoplanetary discs \citep{2013ApJ...769...76B,2014prpl.conf..411T}.
MHD winds are also likely to influence gas-disc migration, though this is an active field of research and the details are not yet clear \citep{Ogihara2015,2021A&A...646A.166L,Paardekooper2023,Wu2025,Hu2025}.
Including the effects of MHD winds in the future would be desirable (see e.g. \citealt{2023A&A...674A.165W,2016A&A...596A..74S}).

Furthermore, it was shown when properly accounting for the interplay of self-gravity and magnetic fields in young discs \citep{2019MNRAS.482.3989R,2020ApJ...891..154D}, GI may also lead to the formation of intermediate-mass planets \citep{2021NatAs...5..440D}.
Accretion and migration in such systems may behave differently in such systems \citep{2023MNRAS.525.2731K,Kubli2025}.

\subsection{Inner disc edge}
We did not study in detail what happens at the inner edge of the disc.
The inner truncation is fixed at \SI{0.05}{au}, and whenever a companion crosses it, it is considered accreted onto the primary.
This assumption is adequate because we are mostly interested in the surviving population at wide orbits.
At the same time, our results show that the number of companions accreted on the central star is comparable to the number of surviving objects.
If even a fraction of the accreted population was able to survive close to the primary, it could explain the formation of some observed companions, for example relatively close giant planets around M-dwarfs \citep{Morales2019}.
A more detailed treatment of the inner edge, including a dependency on the stellar magnetic field and rotation rate would be desirable for this.
Additionally, the companions that do accrete on the primary also need to be considered.
Such events will change the composition of the star and will lead to outbursts that may be observable \citep{2010ApJ...719.1896V,Soares2025}.

\subsection{Metallicity correlations}\label{sec:metal}
Giant planets have long been known to be more frequent around metal-rich stars \citep{1997MNRAS.285..403G,santosisraelian2001}.
This giant planet-metallicity correlation has become well established (e.g. \citealt{Fischer2005}, see \citealt{2019Geosc...9..105A} for a review).
It is considered strong support for CA being responsible for the formation of the majority of giant planets \citep{2004ApJ...604..388I,mordasinialibert2009b,Wang2018,2021A&A...656A..70E}.
There is some discussion in the literature about this trend only being valid up to a ‘transition mass’ of a few \unit{\mj}.
Above this mass, planets and BDs would form through DI \citep{Schlaufman2018,Santos2017a}.
\citet{2019Geosc...9..105A} analyses and discusses these claims in some detail based on a large, homogeneous dataset from the SWEET-Cat catalogue \citep{Santos2013}.
They find that the correlation of planet occurrence with host metallicity continues above \SI{4}{\mj}, though it is less strong.
They conclude that the claim about the different formation mechanism is not justified.
Instead, they argue that the formation environment is different.
This finding seems to be compatible with the results of our study.
As discussed in Sect.~\ref{sec:bn}, our populations seem to produce fewer giant planets between \SIrange{5}{500}{au} than observed by direct imaging.

In its current state, our model would likely predict an anti-correlation of host star metallicity with companion frequency.
This is because in our clump evolution model, clumps of lower metallicity have lower opacities and shorter pre-collapse timescales and are therefore less likely to be tidally or thermally disrupted (see also \citealt{helledbodenheimer2010a,2019MNRAS.488.4873H}).

However, this simplistic argument overlooks important physics.
A higher metallicity could also lead to a lower clump opacity through grain growth and settling.
This is discussed in \citet{helledbodenheimer2011}.
These authors model isolated clumps by solving the stellar structure equations.
They study the effect of grain growth and settling on the pre-collapse timescale of the clumps and find it to be much shorter than without this effect, and insensitive to clump mass.
Additionally, the accretion of pebbles and planetesimals should also be considered \citep{Baehr2019}.
\citet{2015MNRAS.452.1654N} include accretion of pebbles and find a positive correlation of giant planet occurrence with metallicity for planets inside of \SI{5}{au}, and no correlation outside of \SI{10}{au}.
Since our model considers neither accretion of solids nor solid dynamics inside clumps, we cannot make any clear predictions about metallicity correlations at this time.
We plan to address this important topic in a subsequent paper of this series.

\subsection{Clump geometry and clump rotation}\label{sec:rot}
Throughout this work, we assumed that clumps are spherically symmetric objects that do not rotate.
While necessary at the current stage of the model, both assumptions are not satisfied for clumps that form in a disc.
In a more realistic scenario, a clump forming from a collapsing region of a disc will inherit angular momentum from the Keplerian shear and is likely to have an elongated shape.
This is seen clearly in hydrodynamic simulations of disc fragmentation (see e.g. Fig.~2 in \citealt{2010Icar..207..509B}, Fig.~5 in \citealt{2012ApJ...746..110Z}, Fig.~7 in \citealt{2018A&A...618A...7V}, or Fig.~6 in \citealt{Xu2025}).
The clump can also gain angular momentum through gas accretion and will spin up as it contracts \citep{2012ApJ...746..110Z,galvagnihayfield2013}.
Clump rotation could be expected to increase the clump's pre-collapse timescale, as it works against the contraction.
But the effect is more complicated, and the development of non-axisymmetric modes can shorten the pre-collapse \citep{galvagnihayfield2013}.

At this time, it is difficult to judge how these effects will influence the inferred population.
Future studies should include and address these topics.

\section{Summary and conclusions}\label{sec:Conclusions}
\begin{figure*}
  \centering
      \includegraphics[width=0.8\linewidth,frame]{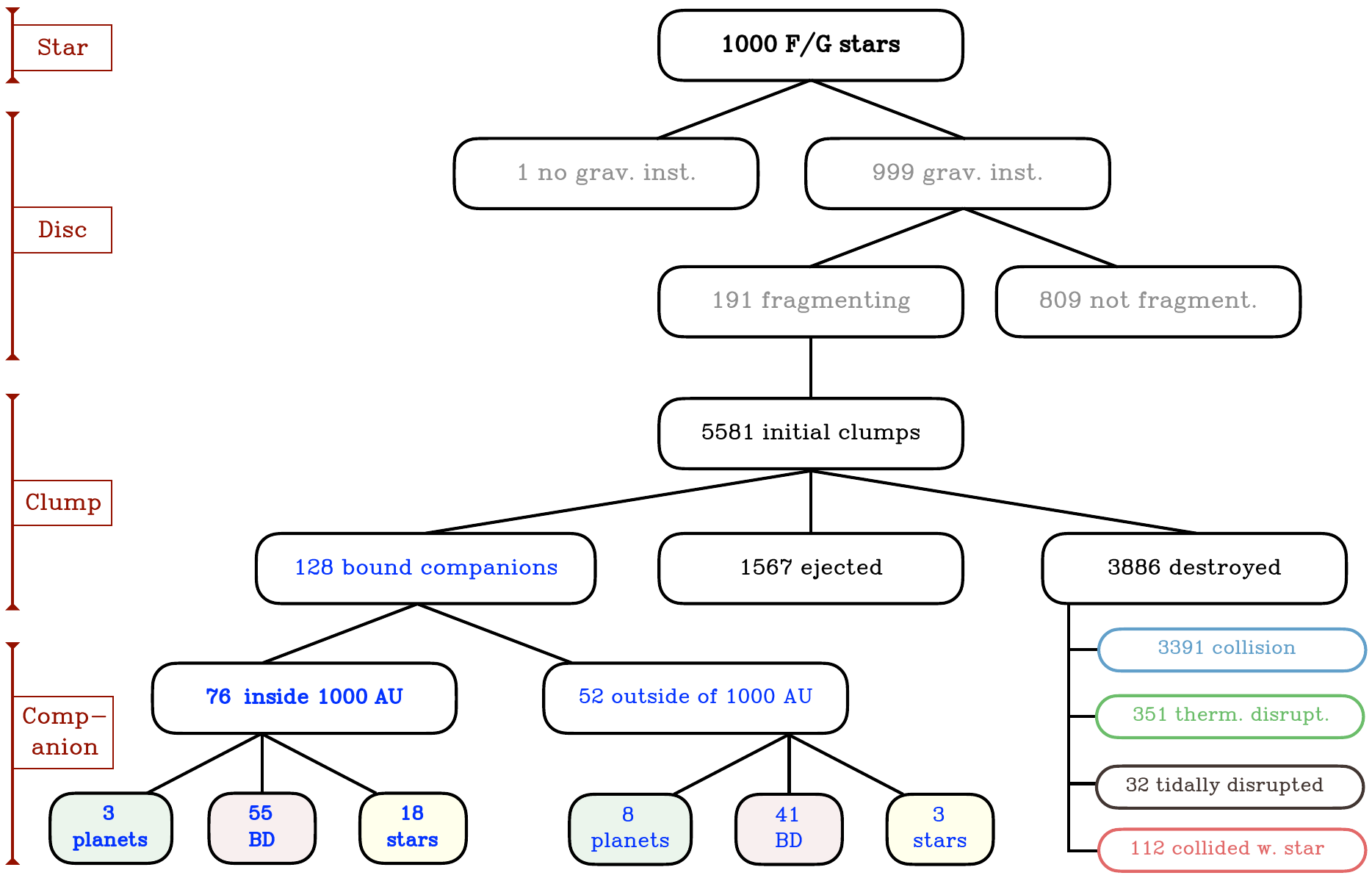}
  \caption{Graphical summary of the results of the baseline population DIPSY\nobreakdash-0 for F/G host stars in terms of the frequency of different outcomes. On the left, the basic level (star, disc, clump, and companion) is shown in brown. The rest shows a decision-tree-like flow chart with the different outcomes. The colour coding is taken from Figs. \ref{fig:am} and \ref{fig:bar_fate} for the mass ranges of the surviving companions and the fates of the destroyed clumps, respectively. A central outcome, the frequency of bound companions inside of 1000 AU, is shown in blue bold face.
  Please note that the numbers are subject to model assumptions.}
  \label{fig:DIPSY-outcome}
\end{figure*}
We performed a large-scale population synthesis study of companion formation in the DI paradigm, applying the model described in \citetalias{Schib2025}.
By combining descriptions of physical processes in star and disc formation and evolution (e.g. infall, disc structure, and stellar evolution), disc GI and fragmentation, multiple clump formation, evolution through gas accretion, interior structure evolution with second collapse, clump orbital migration, clump thermal and tidal destruction, as well as N-body interactions with collisions and many other effects, we were able to globally follow companion formation by DI and quantitatively predicted occurrence rates and physical properties of companions around stars from \SIrange{0.05}{5}{\msun}.
We find DI to be a mechanism that forms companions with masses ranging from the low-mass star down to the planetary regime, although it very rarely produces planetary mass objects inside of about 100 AU (at a per-mill level).
The main results from our baseline population DISPY\nobreakdash-0 are summarised below.
\begin{itemize}
  \item For final stellar masses above 0.1-0.2 $\msun$, virtually all discs undergo an early self-gravitating phase where they form spiral arms.
  \item Fragmentation (i.e. the formation of bound gas clumps) occurs in $\sim$~\SIrange{10}{20}{\%} of systems with final stellar masses $M_{\rm *,final} \gtrsim \SI{0.1}{\msun}$.
  \item When fragmentation occurs, the total number of fragments formed is large, with tens of fragments (clumps) forming on average per fragmenting system and up to hundreds in extreme cases. This high number is explained by the fact that continuing infall drives the discs through several fragmentation events.
  \item However, only a small fraction ($\SI{2}{\%}$) of the clumps survive as bound companions on a \unit{Myr} timescale.
  \item Fragments are lost due to (approximately) collision ($\SI{60}{\%}$), ejection ($\SI{25}{\%}$), thermal disruption from disc irradiation ($\SI{10}{\%}$), accretion on star ($\SI{2}{\%}$), and tidal disruption ($\SI{0.5}{\%}$).
  \item Based on their mass, surviving companions are grouped into low-mass stars (\SI{15}{\%}), BDs (\SI{75}{\%}), and planetary mass objects (\SI{10}{\%}).
  \item The CMF is unimodal and peaks at about \num{50} Jovian masses. Below this value, the frequency of companions scales with mass approximately as $M_\mathrm{c}^{1}$. Our model might underestimate the number of stellar companions formed.
  \item Brown dwarfs and stars occupy a large range of semi-major axes (\SIrange{e1}{e4}{au}), while most planetary mass objects lie beyond \SI{500}{au}. This means that in the mass-distance plane, the envelope of points covers a characteristic inverted and rotated ‘L’ with a paucity of planetary-mass companions inside of 100-1000 AU.
  \item In terms of frequencies, DI produces planetary-mass companions on a low percent level at distances larger than $\sim$\SI{100}{au}. Inside of 100 AU, planetary-mass companions are, in contrast, very rare (low per-mill level).
  \item This key outcome is due to the following physical process based on dynamical clump-clump interactions (scattering) and growth: When clumps remain in the disc, there is typically enough gas for them to grow into the BD or stellar mass regime by both gas accretion \citep{Kratter2010} and clump-clump collisions. Only if they are scattered outwards to large distances beyond the discs' outer radii, and thus deprived of gas to accrete, can they more commonly remain in the planetary mass regime.
  \item Compared to results of direct imaging surveys \citep{2018haex.bookE.155B}, the synthetic occurrence rate of BDs is in good agreement with observations, while the number of planetary mass objects is about a factor of 20 lower, pointing at CA as the dominant formation mechanism \citep{janson2012,Kraus2012}.
  \item Surviving companions have large eccentricities ($e~\approx\num{0.4+-0.2}$) and typical inclinations around \SI{40}{\degree}, both with a large spread.
  \item Surviving companions around more massive stars are on average more massive and on wider orbits, though the dependencies are weak.
  \item Among systems with at least one surviving companion, four out of five only have one single companion; one out of five systems has two; and very few (<~\SI{1}{\%}) have three companions, with almost no primary mass dependency.
  \item Companions are typically $\approx$~\num{25}~$\times$ as massive after \SI{100}{Myr} than at fragmentation, due to gas accretion and mergers.
  \item Clump-clump collisions are very frequent and also contribute considerably to the final masses. We thus find that collisions and gravitational interactions (scatterings) in general are key in shaping the emerging companion population formed by DI. The fact that the outcome is driven by clump-clump interactions as an internal process explains why we observe remarkably weak dependencies of our results on many external (boundary) conditions such as stellar properties but also several important model assumptions.
  \item Ejections, as a related phenomenon, are very frequent and lead to $\gtrsim$~\num{1} ejected object per star for all primaries.
\end{itemize}
Figure \ref{fig:DIPSY-outcome} provides a graphical summary of many of these results, with a flow chart splitting up into different outcomes from the top to the bottom. For a sample of 1000 F/G stars, it first displays the number of discs undergoing GI. Given that we only used a sub-range of (particularly relevant) stellar masses in this study, the exact fractions differ to some extent from the ones given in the textual summary. As discussed, virtually all discs are self-gravitating at some time during the infall and early evolutionary phase. For those undergoing such an instability, we can then see how many fragment (about 19\%).
The majority of systems do not fragment due to the fast transport of mass and angular momentum by the spiral arms (Sect.~\ref{sec:fractions}).
On the next level, the clump level, the total number formed is given, as well as their fundamental fate (ending up as bound companions, ejected, or destroyed).
It is again remarkable to see how many initial clumps form (5581, i.e. on average about 29 per fragmenting disc), but how few (128) remain in the end as bound companions. Ejection as the fate is already more than ten times more common (1567 clumps), but destruction (3886 clumps) is clearly the most likely outcome via one of the four listed mechanisms. Among these, collision with another clump is by far the most important channel. Finally, for the bound companions that remain at the end of the simulation, we classify the bodies according to semi-major axis-- smaller (76) or larger (52) than 1000 au--and further into planetary mass, BD, and stellar companion based on their mass.
We see that inside 1000 AU, only three planetary-mass objects eventually result (about 3\permil), compared to 55 BD and 18 stars.

As a global model, our framework depends on a non-negligible number of parameters and base assumptions about the governing physical processes. We thus also quantified a number of variant populations to study the sensitivity on different parameters and assumptions.
From these we find:
\begin{itemize}
  \item Relinquishing the baseline limit on the gas accretion rate of \num{e-3} Jovian masses per year, often seen in hydrodynamic simulations \citep{2013ApJ...767...63S,2019MNRAS.486.4398F}, can lead to the formation of very massive companions, including equal mass binaries. However, these simulations are not quantitatively physically realistic, as they are not consistent with our base assumption that the central object (primary star) always remains more massive than any companion formed.
  \item Disabling gas accretion altogether and, instead, using a higher initial fragment mass (as in \citealt{2018MNRAS.474.5036F}) leads to a population surprisingly similar to the baseline.
  \item Whether or not clumps have initial eccentricity does not have a significant influence on their survival in our model, a testament to the strong internal driving of the system by gravitational interactions.
  \item A larger, early disc size (larger than observed) in contrast to the calibrated sizes in the baseline population) leads to a disproportionately higher number of fragments, as well as surviving companions. The fraction of planetary mass objects is higher in this case, but the fraction of BDs is too high compared to observations.
  \item The possibility for several fragments to form per fragmentation event increases their chance of survival.
  \item The background alpha-viscosity has a minor influence on the population as long as the disc lifetimes remain similar. The underlying reason is that in the most important early phase, self-gravity determines the actual (high) alpha.
\end{itemize}

We conclude that DI is a robust mechanism for the formation of low-mass stars, BDs and planets, although it largely avoids populating the inner 100-1000 au with planetary mass objects.

The number of planets formed this way may be low (by a factor of $\sim$20 lower when compared to some direct imaging results), in agreement with predictions from the literature \citep{2015MNRAS.454.1940R,Kratter2010}. However, important physics is still missing in our model (and sometimes not even well understood), so it is too early to firmly draw this conclusion (Sect.~\ref{sec:Discussion}). The number of BDs produced by our model agrees well with occurrence rates from direct imaging surveys. A more thorough comparison with observations, including data from upcoming larger surveys, will be conducted in future studies to put the wealth of quantitative predictions provided by the synthetic systems to the observational test. Such comparisons will reveal differences between synthesis and observations; however, the differences will precisely allow us to better understand the formation process and make it possible to improve the models.

In agreement with previous studies \citep{Papaloizou2001,Forgan2015,2018MNRAS.474.5036F}, we predict DI to produce a high number of ejected objects.

We identified a number of processes that have a key influence on the outcome of DI. The most important are the following: the formation of a very high number of clumps due to continuous fragmentation in infalling discs; dynamical interactions (scatterings and ejections) among the clumps concurrently present in a disc; gas accretion by the clumps leading to significant mass growth; and a very high number of clump-clump collisions. The characteristics of disc formation (in particular, the early disc size) are also important. The precise nature of these processes should be studied further by dedicated models in the future.
The results of such studies can then be included in the next generation of population synthesis models.

Some physical mechanisms, such as solid accretion and MHD-effects, are still partly or fully missing from the model \citep{2015MNRAS.452.1654N,Baehr2019,2013ApJ...769...76B,2019MNRAS.482.3989R,2021NatAs...5..440D,2020ApJ...891..154D,2023MNRAS.525.2731K,Kubli2025}.
Including them in the future will be important as they could influence our results in a relevant way.

 Gravitational instabilities (not necessarily including fragmentation) appear to be ubiquitous in protoplanetary discs. Their effect should be considered in studies of planet formation beyond the DI framework as well, as it influences the disc's structure of both the gas and the dust component \citep{Boss2015,2024arXiv240109380R}. Such disc inhomogeneities might play a role in the dynamics of pebbles as well as the formation of planetesimals and of (solid) protoplanets. Disc instability and CA might thus not be considered fully distinct formation pathways. Hybrid models have been proposed (e.g. \citealt{2007prpl.conf..607D}), and they may be necessary to explain some of the recent observations of young protostellar and protoplanetary systems. The quantitative statistical comparison to observations, especially with large surveys such as SHINE \citep{Desidera2021,Langlois2021,2021A&A...651A..72V}, GPIES \citep{Nielsen2019}, BEAST \citep{2021A&A...646A.164J,Delorme2024}, or COBREX \citep{Squicciarini2025}, will help untangle the mysteries of star and planet formation.

\begin{acknowledgements}
We thank Michael Meyer and Lucio Mayer for the insightful discussions.
OS and CM acknowledge the support from the Swiss National Science Foundation under grant \texttt{\detokenize{200021_204847}} 'PlanetsInTime'. R.H. acknowledges support from the Swiss National Science Foundation under grant 200020\_215634. Part of this work has been carried out within the framework of the NCCR PlanetS supported by the Swiss National Science Foundation under grants \texttt{\detokenize{51NF40_182901}} and \texttt{\detokenize{51NF40_205606}}. Calculations were performed on the Horus cluster of the Division of Space Research and Planetary Sciences at the  University of Bern.
\end{acknowledgements}

\bibliographystyle{aa}
\bibliography{main}
\begin{appendix}

\section{Comparison of existing GI population syntheses}
\label{app:comp}

The table in this appendix (adapted from \citet{Schib2021}) gives a  comparison of several  GI population syntheses conducted to date in terms of the physics included in the models.
The different studies use  different prescriptions for migration. The following symbols will be used in the table:
\begin{equation}
t_\mathrm{I,FR} = \left( \frac{M_\mathrm{p}}{\mstar} \Omega \right) ^{-1} \frac{H}{r},
\end{equation}
\begin{equation}
t_\mathrm{I,N} = \frac{\mstar^2}{M_\mathrm{p} M_\mathrm{d,loc}} \Omega_\mathrm{p}^{-1},
\end{equation}
where
\begin{equation}
M_\mathrm{d,loc} = \pi \Sigma r^2,
\end{equation}

\begin{equation}
t_\mathrm{II,FR} = \frac{1}{\alpha \Omega} \left( \frac{H}{r} \right) ^{-2},
\end{equation}
and $M_\mathrm{p}$ is the companion mass.
The updated model from \citet{2018MNRAS.474.5036F} is not included explicitly in the table. It is the only model that includes gravitational interaction between the clumps besides the model presented here. Other than that, the only difference to  \citet{2013MNRAS.432.3168F} is the gap opening criterion given in the explanations below.

Here we give some additional explanations for the following Table~\ref{tab:comp}:\\\\
\textsuperscript{0} The updated model in \citet{2018MNRAS.474.5036F} uses instead the gap opening condition from \citet{2006Icar..181..587C} with an additional condition for gap crossing from \citet{2015ApJ...802...56M}.
\textsuperscript{1} In addition to \citet{2006Icar..181..587C}, the authors investigate two additional conditions related to the gap crossing and gap formation timescale.  \\
\textsuperscript{2} with $\alpha = 0.005$\\
\textsuperscript{3} fragments are randomly space within the unstable region\\
\textsuperscript{4} disc is unstable at the beginning of the simulation\\
\textsuperscript{5} fragment is inserted at the initial fragment location\\
\textsuperscript{6} \citet{2007MNRAS.375..500A} for UV and \citet{2012MNRAS.422.1880O} for X-ray\\
\textsuperscript{7} External photoevaporation as in \citet{2007MNRAS.376.1350C}\\
\textsuperscript{8} All models use the `first core/follow the adiabats' approach described in Sect.~4 of \citet{2015MNRAS.454...64N}.\\
\textsuperscript{9} An N-body integrator is used in the updated model in \citet{2018MNRAS.474.5036F}.

\section{Infall locations}\label{app:rinf}
Table \ref{table:infrad} lists the infall location $R_\mathrm{i}$ of the baseline population DIPSY\nobreakdash-1 for each of the \num{100} logarithmically spaced bins.
The bins are for the final stellar mass $M_\mathrm{*,final}$.
\begin{table}[ht]
\caption{Infall locations for the baseline population.}
\begin{tabular}{lll|lll}
\hline\hline
\begin{tabular}[c]{@{}c@{}}\bf{bin}\\\bf{index}\end{tabular} &
\begin{tabular}[c]{@{}c@{}}\bf{centre}\\\bf{(\unit{\msun})}\end{tabular} &
\begin{tabular}[c]{@{}c@{}}\bf{$R_\mathrm{i}$}\\\bf{(\unit{au})}\end{tabular} &
\begin{tabular}[c]{@{}c@{}}\bf{bin}\\\bf{index}\end{tabular} &
\begin{tabular}[c]{@{}c@{}}\bf{centre}\\\bf{(\unit{\msun})}\end{tabular} &
\begin{tabular}[c]{@{}c@{}}\bf{$R_\mathrm{i}$}\\\bf{(\unit{au})}\end{tabular} \\
\hline
1   & 0.051 & 16.2    & 51  & 0.512 & 11.9    \\
2   & 0.054 & 15.7    & 52  & 0.536 & 12.0    \\
3   & 0.056 & 17.5    & 53  & 0.561 & 12.1    \\
4   & 0.059 & 13.7    & 54  & 0.587 & 12.2    \\
5   & 0.062 & 17.5    & 55  & 0.615 & 12.3    \\
6   & 0.064 & 19.3    & 56  & 0.644 & 12.3    \\
7   & 0.067 & 14.3    & 57  & 0.674 & 12.4    \\
8   & 0.071 & 15.7    & 58  & 0.706 & 12.5    \\
9   & 0.074 & 14.6    & 59  & 0.740 & 12.6    \\
10  & 0.077 & 15.8    & 60  & 0.774 & 12.7    \\
11  & 0.081 & 17.2    & 61  & 0.811 & 12.8    \\
12  & 0.085 & 15.3    & 62  & 0.849 & 12.9    \\
13  & 0.089 & 17.6    & 63  & 0.889 & 13.0    \\
14  & 0.093 & 16.4    & 64  & 0.931 & 13.1    \\
15  & 0.097 & 14.6    & 65  & 0.975 & 13.1    \\
16  & 0.102 & 13.2    & 66  & 1.021 & 13.2    \\
17  & 0.107 & 11.0    & 67  & 1.069 & 13.3    \\
18  & 0.112 & 10.3    & 68  & 1.119 & 13.4    \\
19  & 0.117 & 10.3    & 69  & 1.172 & 13.4    \\
20  & 0.123 & 9.8     & 70  & 1.227 & 13.5    \\
21  & 0.129 & 9.5     & 71  & 1.285 & 13.6    \\
22  & 0.135 & 9.6     & 72  & 1.346 & 13.7    \\
23  & 0.141 & 9.9     & 73  & 1.409 & 13.7    \\
24  & 0.148 & 9.2     & 74  & 1.476 & 13.8    \\
25  & 0.155 & 9.7     & 75  & 1.545 & 13.9    \\
26  & 0.162 & 10.1    & 76  & 1.618 & 13.9    \\
27  & 0.169 & 9.2     & 77  & 1.694 & 14.0    \\
28  & 0.177 & 10.0    & 78  & 1.774 & 14.0    \\
29  & 0.186 & 10.2    & 79  & 1.858 & 14.1    \\
30  & 0.195 & 10.1    & 80  & 1.945 & 14.1    \\
31  & 0.204 & 10.4    & 81  & 2.037 & 14.2    \\
32  & 0.213 & 10.5    & 82  & 2.133 & 14.2    \\
33  & 0.223 & 10.2    & 83  & 2.233 & 14.3    \\
34  & 0.234 & 10.5    & 84  & 2.339 & 14.3    \\
35  & 0.245 & 10.4    & 85  & 2.449 & 14.4    \\
36  & 0.256 & 10.9    & 86  & 2.564 & 14.4    \\
37  & 0.269 & 10.6    & 87  & 2.685 & 14.4    \\
38  & 0.281 & 11.4    & 88  & 2.812 & 14.5    \\
39  & 0.294 & 10.5    & 89  & 2.944 & 14.5    \\
40  & 0.308 & 10.8    & 90  & 3.083 & 14.6    \\
41  & 0.323 & 11.0    & 91  & 3.228 & 14.6    \\
42  & 0.338 & 11.2    & 92  & 3.380 & 14.6    \\
43  & 0.354 & 11.0    & 93  & 3.540 & 14.6    \\
44  & 0.371 & 11.2    & 94  & 3.707 & 14.7    \\
45  & 0.388 & 11.3    & 95  & 3.881 & 14.7    \\
46  & 0.406 & 11.4    & 96  & 4.064 & 14.7    \\
47  & 0.426 & 11.5    & 97  & 4.256 & 14.8    \\
48  & 0.446 & 11.6    & 98  & 4.456 & 14.8    \\
49  & 0.467 & 11.7    & 99  & 4.666 & 14.8    \\
50  & 0.489 & 11.8    & 100 & 4.886 & 14.8    \\
\hline
\end{tabular}
\tablefoot{Bin index, bin centre, and infall location ($R_i$) for DIPSY\nobreakdash-0.}
\label{table:infrad}
\end{table}

\thispagestyle{empty}
{\addtolength\textheight{1cm}
\begin{landscape}
\begin{table}[]
\caption{Comparison of existing GI population syntheses.}
\centering
\resizebox{25cm}{!}{
\begin{tabular}{llllllll}
\hline\hline
Publication                                       & \citet{2013MNRAS.432.3168F}   & \citet{2018ApJ...854..112M}& \citet{2015MNRAS.454...64N}   & \citet{2015arXiv150207585N}            & \citet{2015MNRAS.452.1654N} & \citet{2016MNRAS.461.3194N} & \citet{2019MNRAS.488.4873H}  \\
\hline
\begin{tabular}[c]{@{}l@{}}Title\end{tabular} &
\begin{tabular}[t]{@{}l@{}}Towards a population\\ synthesis model of\\ objects formed by\\ self-gravitating\\ disc fragmentation\\ and tidal downsizing\end{tabular} &
\begin{tabular}[t]{@{}l@{}}On the Diversity in\\ Mass and Orbital\\ Radius of Giant\\ Planets Formed via\\ Disk Instability\end{tabular} &
\begin{tabular}[t]{@{}l@{}}Tidal Downsizing\\ model - I. Nu-\\merical methods:\\ saving giant pla-\\nets from tidal\\ disruptions\end{tabular} &
\begin{tabular}[t]{@{}l@{}}Tidal Downsizing\\ model. II. Planet-\\metallicity\\ correlations\end{tabular} &
\begin{tabular}[t]{@{}l@{}}Tidal downsizing\\ model - III.\\ planets from sub-\\earths to brown\\ dwarfs: Structure\\ and metallicity\\ preferences\end{tabular} &
\begin{tabular}[t]{@{}l@{}}Tidal Downsizing\\ model - IV.\\ Destructive feed-\\ back in planets\end{tabular} &
\begin{tabular}[t]{@{}l@{}}Constraining the\\ initial planetary\\ population in the\\ gravitational insta-\\ bility model\end{tabular} \\
\hline
Viscous $\alpha$                            & 0.005                         & 0.005                      & 0.01                          & [0.001..0.01]                          & 0.005                            & [0.005..0.05]          & 0.005                    \\
GI $\alpha$                                 & \citet{2001ApJ...553..174G}   & \citet{2008ApJ...681..375K}& ---                           & $0.2 \frac{Q_0^2}{Q_0^2+Q^2}, Q_0 = 2$ & $0.2 \frac{Q_0^2}{Q_0^2+Q^2}, Q_0 = 5$ & $0.2 \frac{Q_0^2}{Q_0^2+Q^2}, Q_0 = 1.5$& 0.12    \\
Type I mig                                  & $t_\mathrm{I,FR}$                    & \citet{2011MNRAS.416.1971B}& $t_\mathrm{I,N} \times \left( 1 + \frac{M_\mathrm{p}}{M_\mathrm{d}} \right)$
                                                                                                                                         & $[1..10] \times t_\mathrm{I,N}$               & $[0.5..2] \times t_\mathrm{I,N}$        & $[1..4] \times t_\mathrm{I,N}$& $0.8 \times t_\mathrm{I,N}$     \\
Type II mig                                 & $t_\mathrm{II,FR}$                   & viscous                    & impulse approx.               & impulse approx.                        & impulse approx.                  & impulse approx.        & $t_\mathrm{II,FR} \times \left( 1 + \frac{M_\mathrm{p}}{M_\mathrm{d}} \right)$  \\
Gap opening                                 & \citet{2002MNRAS.330L..11A}\textsuperscript{0}
                                                                                                         & various\textsuperscript{1} & \citet{2006Icar..181..587C}   & \citet{2006Icar..181..587C}            & \citet{2006Icar..181..587C}      & \citet{2006Icar..181..587C}& \citet{2006Icar..181..587C}\textsuperscript{2} \\
Gas accretion                               & ---                           & Galvani \& Mayer (2014)    & ---                           & ---                                    & ---                              & ---                    & ---                      \\
Pebble accretion                            & ---                           & ---                        & \Haek                         & \Haek                                  & \Haek                            & \Haek                  & ---                      \\
\# fragments                                & several\textsuperscript{3}    & 1                          & 1                             & 1                                      & 1                                & 1                      & 1                        \\
Condition for frag.                         & \citet{2011MNRAS.417.1928F}   & ---\textsuperscript{4}     & ---\textsuperscript{5}        & ---\textsuperscript{5}                 & ---\textsuperscript{5}           & ---\textsuperscript{5} & ---\textsuperscript{5}   \\
2nd collapse                                & \Haek                         & \Haek                      & \Haek                         & \Haek                                  & \Haek                            & \Haek                  & \Haek                    \\
Initial fragment location                   & randomly in unstable region   & [80..120] \si{au}          & \SI{100}{au}                  & [60..120] \si{au}                      & [70..105] \si{au}                & [70..105] \si{au}      & [40..300] \si{au}        \\
Tidal downsizing                            & \Haek                         & \Haek                      & \Haek                         & \Haek                                  & \Haek                            & \Haek                  & \Haek                    \\
Thermal bath disruption                     & ---                           & ---                        & \Haek                         & \Haek                                  & \Haek                            & \Haek                  & \Haek                    \\
Initial stellar mass                        & [0.8..1.2] \si{\msun}         & 0.6, 1.0, 1.4, 2.4 \si{msun}& \SI{1}{\msun}                & \SI{1}{\msun}                          &  \SI{1}{\msun}                   & \SI{1}{\msun}          & \SI{1}{\msun}            \\
Initial disc mass                           & [0.125..0.375] \si{\msun}     & 0.18, 0.3, 0.42, 0.72 \si{\msun}& \SI{0.12}{\msun}         & [0.05..0.2] \si{\msun}                 & [0.075..0.15] \si{\msun}         & [0.075..0.2]\si{\msun} & [0.1..0.3] \si{\msun}    \\
Initial $q = M_\mathrm{d}/\mstar$                    & constant                      & 0.3                        & 0.12                          & 0.05-0.2                               & 0.075-0.15                       & 0.075-0.2              & 0.1-0.3                  \\
Initial disc radius                         & [50..100] \si{au}             & $\approx$\SI{120}{au}         & \SI{120}{au}                  & \SI{100}{au}                           & \SI{100}{au}                     & \SI{100}{au}           & 80-300 au                \\
Initial surface density profile             & -1                            & -1/2, -1, -3/2             & -1                            & -1                                     & -1                               & -1                     & -1                       \\
Photoevaoration                             & \citet{2010MNRAS.401.1415O}   & ---                        & ---                           & UV and X-ray\textsuperscript{6}        & \textsuperscript{6} and \textsuperscript{7} & \textsuperscript{6} and \textsuperscript{7}&---\\
Initial fragment mass                       & \citet{2011MNRAS.417.1928F}   & \SIrange{0.5}{16}{\mj}     & \SI{1}{\mj} or \SI{5}{\mj}    & 0.5, 0.7, 1 or 2 \si{\mj}              & [0.333..16]\si{\mj}              & [0.333..8] \si{\mj}    & 1, 3, 5, 7 \si{\mj}      \\
Clump structure                             & FC\textsuperscript{8}         & varied                     & FC\textsuperscript{8}         & FC\textsuperscript{8}                  & FC\textsuperscript{8}            & FC\textsuperscript{8}  & Vazan \& Helled (2012)   \\
Clump contraction                           & \Haek                         & \Haek                      & \Haek                         & \Haek                                  & \Haek                            & \Haek                  & \Haek                    \\
Grain growth                                & \Haek                         & ---                        & \Haek                         & \Haek                                  & \Haek                            & \Haek                  & ---                      \\
Grain sedimentation                         & \Haek                         & ---                        & \Haek                         & \Haek                                  & \Haek                            & \Haek                  & ---                      \\
Core formation and growth                   & \Haek                         & ---                        & \Haek                         & \Haek                                  & \Haek                            & \Haek                  & ---                      \\
N-body interaction                          & --- / (\Haek)\textsuperscript{9}& ---                      & ---                           & ---                                    & ---                              & ---                    & ---                      \\
\hline
\end{tabular}}
\tablefoot{If ranges are given in brackets, it means the corresponding values are sampled in a Monte Carlo approach from the given interval. The updated Forgan \& Rice mode, 'Towards a population synthesis model of self-gravitating disc fragmentation and tidal downsizing II: the effect of fragment-fragment interactions', is not included in a separate row. However, the differences are explained. See text for further explanations.}
\label{tab:comp}
\end{table}
\end{landscape}
\addtolength\textheight{0cm}
\clearpage
}

\end{appendix}

\end{document}